\documentclass[british,12pt,a4paper]{article}
\usepackage{a4wide}

\usepackage[utf8]{inputenc}
\usepackage[T1]{fontenc}
\usepackage{babel}
\usepackage[autostyle=true]{csquotes}
\usepackage{amsmath}
\usepackage{bm}
\usepackage{amssymb}
\usepackage{mathtools}
\usepackage{physics}
\usepackage{graphicx}
\usepackage{float}
\usepackage{caption}
\usepackage{subcaption}
\usepackage{multirow}
\usepackage{appendix}
\usepackage{csquotes}
\usepackage[colorlinks,linkcolor=Blue, urlcolor=Blue,citecolor=BrickRed]{hyperref}
\usepackage{typearea}
\usepackage{cancel}
\usepackage{cleveref}
\usepackage{braket}
\usepackage{placeins}
\usepackage[dvipsnames]{xcolor}
\usepackage{simpler-wick}
\usepackage{tikz}
\usetikzlibrary{decorations.markings}
\usepackage{soul}
\usepackage{array}
\usepackage{physics}
\usepackage[nosort,noadjust]{cite}

\crefname{appsec}{appendix}{appendecies}
\Crefname{appsec}{Appendix}{Appendecies}

\numberwithin{equation}{section}
\numberwithin{figure}{section}

\newcommand{\RNum}[1]{\uppercase\expandafter{\romannumeral #1\relax}}
\DeclareMathOperator{\vecspan}{span}

\DeclareMathOperator\artanh{artanh}

\title{Grand Canonical vs Canonical Krylov Complexity in Double-Scaled Complex SYK Model}
\author{Stefan F\"orste, Yannic Kruse, Saurabh Natu}
\date{June 2025}

\begin{document}

\thispagestyle{empty}
\vspace*{-2.5cm}
\begin{center}
\hfill BONN--TH--2025--35
\end{center}
\vskip 0.6in
\begin{center}
{\bf \Large Grand Canonical vs Canonical Krylov Complexity in Double-Scaled Complex SYK Model}
\end{center}  

\vspace*{1cm}

\centerline{Stefan F{\"o}rste$^a$, Yannic Kruse$^a$, Saurabh Natu$^a$ }
\vskip 0.25in
\begin{center}{\it $^a$ Bethe Center for Theoretical Physics, Physikalisches Institut der Universität Bonn,\\
Nussallee 12, D-53115 Bonn, Germany}
\vskip 0.15in
{\tt forste@th.physik.uni-bonn.de}\\
{\tt ykruse@uni-bonn.de}\\
{\tt snatu@uni-bonn.de}
\end{center}
\vskip 0.15in
\begin{abstract}
	We consider the complex SYK model in the double-scaling limit.
	We obtain the transfer matrix for the grand canonical ensemble and symmetrize it.
	In the $(n, Q)$-basis of chord states, the grand canonical transfer matrix is block diagonal, where each block is the canonical transfer matrix for the respective charge sector.
	We therefore conclude that the Krylov complexity for the grand canonical ensemble is given by the sum of the complexities in the charge sectors weighted by a probability function that depends on the chemical potential.
	Finally, we compute the Krylov complexity analytically in the limit of early and late time in the charge sector and numerically for both canonical and grand canonical ensemble. 
\end{abstract}
	
	\newpage
	\tableofcontents
	\newpage

	\section{Introduction}
		\label{sec:intro}
The ideas of holography \cite{tHooft:1993dmi,Susskind:1994vu} in particular the AdS/CFT correspondence \cite{Maldacena:1997re} have provided a multitude of methods to investigate the features of black holes.
One such feature is the volume of the Einstein-Rosen Bridge (ERB) as a function of time \cite{Brown:2015lvg}.
Before scrambling time, the ERB volume increases exponentially fast. 
While after, it enters a regime of linear growth until the Heisenberg time at which it reaches its maximum.
The behaviour of Krylov complexity of operators and states defined via notions of quantum information theory \cite{Parker:2018yvk, Barbon:2019wsy, Rabinovici:2020ryf, Balasubramanian:2022tpr} has a similar behaviour \cite{Susskind:2014Comp, Susskind:2014rva, Nandy:2024evd,Rabinovici:2025otw}.

For two-dimensional Jackiw-Teitelboim (JT) gravity 
\cite{Jackiw:1984je, Teitelboim:1983ux}, the volume of the 
ERB at a given point in boundary time is replaced by its 
renormalized length in the bulk \cite{Harlow:2018tqv}.
The duality between JT gravity and the Sachdev-Ye-Kitaev (SYK) 
model 
\cite{Sachdev:1992fk, KitaevEntangled2015, 
Maldacena:2016hyu, Kitaev:2017awl} provides a way to 
map the renormalized length of the ERB in the bulk to an operator in the boundary theory.
Such a relation was obtained in the double-scaling limit of 
the SYK model \cite{Cotler:2016fpe, Berkooz:2018qkz, 
Berkooz:2018jqr}.
The chord diagram formalism was adapted to compute the partition
function as well as the correlation functions of boundary operators in
the double-scaling limit of the complex SYK model
\cite{Berkooz:2020uly,Narayan:2023wlk} and the $\mathcal{N}=1,2$ supersymmetric SYK
models \cite{Fu:2016vas,Berkooz:SUSY_SYK}.  

For the $\mathcal{N}=2$ case, it was shown in \cite{Lin:2022zxd} that a two-sided wormhole Hilbert
space appears in the computation of a one-sided thermal
partition function. 
Using this fact, the author of \cite{Lin:2022rbf} described how the
same two-sided wormhole Hilbert space could be obtained by expressing
the partition function in terms of an open chord basis. 
The Hamiltonian acting on the one-sided thermal states is the quantum
mechanical dual given by the SYK model, while the Hamiltonian acting
on the two-sided wormhole states is the Liouville Hamiltonian
\cite{Bagrets:2016cdf, Harlow:2018tqv}. 
The renormalized length $l$ of these two-sided wormholes was 
associated to the chord number operator
defined on the boundary theory for the normalized chord basis. 
By re-expressing the partition function in terms of the chord basis, the
effective Hamiltonian of the SYK model could 
be related to a Hermitian transfer matrix.
The algebra of the creation and annihilation operators involved in the
definition of the Hermitian transfer matrix along with the chord
number operator were seen to form an algebra of the q-deformed
oscillator \cite{Arik:1973vg}. 
Moreover, rather than simply considering empty wormholes, boundary
operators can be introduced.
The resulting correlation functions as
well as the algebra was analysed in \cite{Lin:2022rbf,Lin:2023trc}. 
Using these results the authors of \cite{Ambrosini:2024sre} were able
to compute the 
state and operator Krylov complexity in both the double- and the
triple-scaling limit of the SYK model. 
The Krylov complexity for deformed versions of Majorana SYK, complex SYK
and the $\mathcal{N} = 2$ case have been studied using the Lanczos
algorithm in \cite{Chapman:2024pdw,Aguilar-Gutierrez:2025sqh,Chryssanthacopoulos:2025xyn}.

In a recent paper, the authors of \cite{Caputa:2025mii} discuss the
Krylov complexity of $U(1)$-symmetric systems. 
A major focus of that paper is the relation between the Krylov
complexity in the grand canonical ensemble,  
$C_\mathcal{K}(t)$,
and the canonical ensemble, $C_\mathcal{K}^{(Q)}(t)$, in a sector with
fixed $U(1)$-charge, $Q$.  
They postulate that the grand canonical Krylov complexity of some operator $O$,
with $[O,Q]=0$, 
is greater than the weighted sum of the complexities in the charge sectors, 
$\overline{C}(t)=\sum p_Q C_\mathcal{K}^{(Q)}(t)$, at all times,
\begin{align}
    C_\mathcal{K}(t)-\overline{C}(t) \ge 0, \label{eq:krylov_inequality}
\end{align}
with equality iff the corresponding grand canonical transfer matrix
does not mix sectors. 

In the present paper, we consider the $U(1)$-symmetric complex SYK
model (cSYK) \cite{Berkooz:2020uly}. 
The aim of this paper is twofold.
First, we wish to build up the machinery necessary to study cSYK correlation functions with charged operator insertions.
For this endeavour the oriented chord formulation seems indispensable.
We largely explain it in \cref{sec:moments}.
To get the grand canonical transfer matrix, we first obtain the  $H$-chord version of the transfer matrix for the canonical ensemble and diagonalize it (\cref{sec:fixedQTransferMatrix}).
We then find the symmetrized grand canonical transfer matrix in the oriented chord formulation through comparison with the canonical transfer matrix (\cref{sec:Compare_To_GC}).
As we are currently aware, a symmetrized transfer matrix for the cSYK has never been formulated neither in terms of oriented nor unoriented chords.
The grand canonical transfer matrix will be necessary to study the bulk picture.
The dual for the cSYK has been studied extensively in the low energy limit \cite{Gaikwad:2018dfc,Mertens:2019tcm,Gubankova:2025gbx}.
We expect that a holographic dual akin to the one studied in \cite{Blommaert:2024ymv} for the DSSYK model exists for the full double-scaled cSYK as well.
We leave that for future work, where we will also give a bulk interpretation of our findings for the Krylov complexity.

Second, we wish to study the inequality \eqref{eq:krylov_inequality} conjectured in \cite{Caputa:2025mii}.
We show that in the case of the cSYK the grand canonical
transfer matrix is just a sum of transfer matrices in the charge
sectors,  
i.e.\ it does not mix charge sectors, and equality in
(\ref{eq:krylov_inequality}) is realized. 
In \cref{sec:Krylov}, we compute the spread complexity,  
i.e.\ the Krylov complexity of the number operator, for both the canonical and grand canonical ensemble numerically and compare it with analytically obtained asymptotes for very early and very late times.

\vspace*{0.5in}

\noindent {\bf Note added:} While finalizing the present paper, 
\cite{Gubankova:2025gbx} appeared on the arXiv. 
That paper also 
considers the grand canonical moments albeit using an 
alternative approach.
	\section{cSYK Moments with Benefits}
		\label{sec:moments}
In this section, we introduce some useful machinery necessary to
obtain the transfer matrices in the grand canonical and canonical
ensembles in the next section. 
To do so, in \cref{subsec:cSYK_setup} we provide the general setup
for taking the double-scaling limit of the cSYK model. 
Next, in \cref{subsec:GC_oriented_chords} we recast the moments of the
grand canonical partition function in terms of oriented chords
(following the procedure described in \cite{Berkooz:SUSY_SYK}). 
In \cref{subsec:Setting_up_T}, we obtain the penalty factors for
closing chords. 
Further, in \cref{sec:m0sector} the inner product (defined
in \cref{sec:inner_product}) is employed
along with the saddle point approximation as described in
\cite{Berkooz:2020uly} to provide an explanation for restricting to
the sector with an equal number of oppositely oriented open chords ($m=0$) and
obtain the penalty factors in the charged sectors. 
Finally, in \cref{sec:bosonic_sector} we use the inner product to obtain the associated algebra of
normalized $H$-chord creation and annihilation operators acting on
normalized grand canonical as well as canonical states.  
\subsection{The Complex SYK Model}\label{subsec:cSYK_setup}
We obtain the $U(1)$-symmetric cSYK model from the real SYK, by
replacing the Majorana fermions with Dirac fermions. 
As is customary in the double scaling limit, we make use of index set notation where, $I$ denotes a 
multi-index, $I = \{i_1,i_2,\cdots,i_p\}$ of distinct $p$ indices $i_1<i_2<\cdots <i_p$ and $\psi_I = \psi_{i_1}\psi_{i_2}\cdots\psi_{i_p}$
\footnote{For simplicity, we follow the same notation defining the
index sets as in \cite{Berkooz:2020uly}, namely here $p = q/2$ in
\cite{Davison:2016ngz}.}. 
The corresponding Hamiltonian reads as follows
\cite{Davison:2016ngz,Bhattacharya:2017vaz,Gu:2019jub} 
\begin{align}
    H_\text{cSYK}=  \sum_{I,I'} J_I^{I'} \, \bar{\psi}_{I'} \psi^{I},
\end{align}
and the random couplings $J_I^{I'}$ are drawn from a Gaussian distribution,
\begin{align}
		&\overline{J_I^{I'}} = 0,&
		\overline{\abs{J_I^{I'}}^2} &= \frac{\mathcal{J}^2}{\lambda} \begin{pmatrix}
        N\\ p
        \end{pmatrix}^{-2}\label{mean_variance_cSYK},
\end{align}
with
\begin{align}
    J_{\dots ij \dots;\dots kl\dots} = - J_{\dots ji\dots;\dots kl\dots} = - J_{\dots ij\dots;\dots lk\dots } = J_{\dots kl\dots ;\dots ij\dots}^*
\end{align}
and
\begin{equation}
\lambda = \frac{p^2}{N}.
\end{equation}
The double scaling limit is defined by taking $N,p \to \infty$ while $\lambda$ is kept finite. 
For large $p$, usual ($N\to \infty$) cSYK results can be
matched with the triple scaling limit in which also $\lambda$ is taken to zero.
The $\psi_i$, $\bar{\psi}_j$ are normalized fermionic operators. 
Their canonical anti-commutators are
\begin{align}
    &\poissonbracket{\psi_i}{\bar{\psi}_j} = 2\delta_{ij}, &\poissonbracket{\psi_i}{\psi_j} = \poissonbracket{\bar{\psi}_i}{\bar{\psi}_j} = 0. \label{eq:clifford}
\end{align}
\subsection{Grand Canonical Moments and Oriented Chord Diagrams}\label{subsec:GC_oriented_chords}
The grand canonical partition function for the complex SYK model is given by
\begin{equation}\label{eq:GC_partition_function}
	\mathcal{Z}(\mu,\beta) = \left<\Tr e^{-\beta K}\right>_J,
\end{equation}
where $K$ is the sum of the Hamiltonian and the $U(1)$-charge operator times the chemical potential\footnote{
Notice the different scaling of this term compared to \cite{Davison:2016ngz}. 
We will use a rescaled version of $\mu$ in \cref{sec:Krylov} and \cref{sec:saddle_appendix} to get the same results for thermodynamic quantities as \cite{Davison:2016ngz} later on.
For now, we use the scaling of \cite{Berkooz:2020uly}} $\mu$,
\begin{equation}
	K = \displaystyle\sum_{I,I'}J^{I'}_I\Bar{\psi}_{I'}\psi^I +
	\frac{\mu}{2\beta}\sum_{i=1}^N\left(\Bar{\psi}_i\psi^i-\psi^i\Bar{\psi}_i\right).\label{eq:K} 
\end{equation}
Expanding \cref{eq:GC_partition_function} in terms of the moments we obtain,
\begin{equation}
	\mathcal{Z}(\mu,\beta) = \sum_{k=0}^\infty\frac{(-\beta)^k}{k!}m_k(\mu),
\end{equation}
with $m_k(\mu)$ defined as,
\begin{align}
	m_k(\mu) &= \left<\Tr H^k \exp\left(-\frac{\mu}{2}\sum_{i=1}^N\left(\Bar{\psi}_i\psi^i-\psi^i\Bar{\psi}_i\right)\right)\right>_J\\
	&= \frac{\mathcal{J}^k}{\lambda^\frac{k}{2}}\sum_{\text{OCD}}\begin{pmatrix}N\\p\end{pmatrix}^{-k}\!\!\!
	\sum_{I_1,\cdots ,I_k}\!\!\!\Tr\!\left[\Bar{\psi}_{I_1}\psi^{I_2}\!\!\!\ldots\! \Bar{\psi}_{I_2}\psi^{I_1}
	\!\!\!\ldots\! \exp\!\!\left( \! -\frac{\mu}{2}\sum_{i=1}^N\left(\Bar{\psi}_i\psi^i-\psi^i\Bar{\psi}_i\right)\!\!\!\right)
	\right]\label{eq:GC_moments}.
\end{align}
In \cref{eq:GC_moments}, the trace over the product of $\psi$ and $\Bar{\psi}$ is evaluated using the method described in \cite{Berkooz:2020uly}. 
The various types of resulting chord diagrams can be grouped into two types depending on whether two multi-indices $I_i$ and $I_j$ 
share a set of indices or not, i.e.\ depending on if $\abs{I_i\cap I_j} \equiv p_{ij}\neq 0$.
From this, we get the various weights for the possible chord diagrams
as tabulated in \cref{fig:cSYK_chord_diagram_weights}. 
Note that here we left implicit that in our case the sum over chord diagrams is restricted to pairwise contractions of $\psi$s and $\bar{\psi}$s 
associated with the same Hamiltonian insertion.
We chose to ignore this for now and enforce this behaviour later on
when we employ the $H$-chord formalism. 
This will keep our analysis more general, so that it can be easily adapted to similar models such as the SUSY SYK.
Moreover, it will serve to show clearly why in this model we do not find correlations between different charge sectors.

From \cite{Berkooz:SUSY_SYK}, we know that diagrams such as $(b),(d),(f)$ in \cref{fig:cSYK_chord_diagram_weights} are non-local 
i.e.\ we require the knowledge of the previously closed chords to account for the diagram properly when disentangling the chord diagrams. 
To do so, we adapt the procedure described in \cite{Berkooz:SUSY_SYK} and apply it to the case of the cSYK model in the double scaling limit. 
Following this procedure, we obtain an expression for the local transfer matrix. 
\begin{table}[h!]
    \centering
    \renewcommand{\arraystretch}{1.5} 
    \begin{tabular}{>{\centering\arraybackslash}m{0.35\textwidth}
                    >{\centering\arraybackslash}m{0.35\textwidth}
                    >{\centering\arraybackslash}m{0.3\textwidth}}
    \multicolumn{2}{c}{\textbf{Oriented chord diagrams}} & \textbf{Weights} \\ \hline
    
    \begin{subfigure}[b]{\linewidth}
        \includegraphics[width=\linewidth]{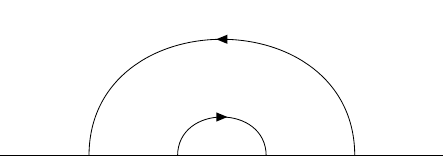}
        \caption{}
    \end{subfigure}
    &\vspace{.72cm}
    \begin{subfigure}[h]{\linewidth}
        \includegraphics[width=\linewidth]{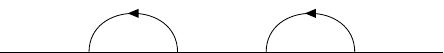}
        \caption{}
    \end{subfigure}
    &
    $\displaystyle \exp[\lambda(2\text{e}^\mu\cosh\mu-1)]$ \\
    
    \begin{subfigure}[b]{\linewidth}
        \includegraphics[width=\linewidth]{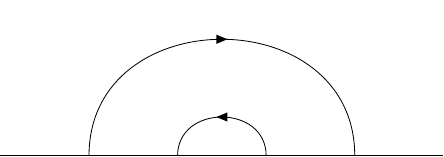}
        \caption{}
    \end{subfigure}
    &\vspace{.72cm}
    \begin{subfigure}[h]{\linewidth}
        \includegraphics[width=\linewidth]{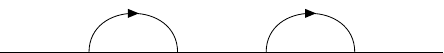}
        \caption{}
    \end{subfigure}
    &
    $\displaystyle \exp[\lambda(2\text{e}^{-\mu}\cosh\mu-1)]$ \\
    
    \begin{subfigure}[b]{\linewidth}
        \includegraphics[width=\linewidth]{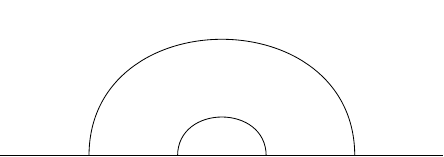}
        \caption{}
    \end{subfigure}
    &\vspace{.72cm}
    \begin{subfigure}[h]{\linewidth}
        \includegraphics[width=\linewidth]{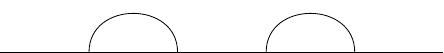}
        \caption{}
    \end{subfigure}
    &
    $\displaystyle \exp(-\lambda)$ \\
    
    \begin{subfigure}[b]{\linewidth}
        \includegraphics[width=\linewidth]{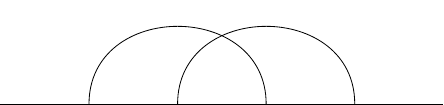}
        \caption{}
    \end{subfigure}
    &
    \hfill
    &
    $\displaystyle (-1)^{p^2}\exp(-\lambda)$ \\
    
    \begin{subfigure}[b]{\linewidth}
        \includegraphics[width=\linewidth]{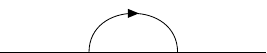}
        \caption{}
    \end{subfigure}
    &
    \hfill
    &
    $\displaystyle \left[\frac{e^\mu}{\cosh\mu}\right]^p$ \\
    
    \begin{subfigure}[b]{\linewidth}
        \includegraphics[width=\linewidth]{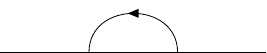}
        \caption{}
    \end{subfigure}
    &
    \hfill
    &
    $\displaystyle \left[\frac{e^{-\mu}}{\cosh\mu}\right]^p$ \\
    
    \end{tabular}
    \caption{Weights for the various chord diagrams. 
	The pair of chords in all diagrams except $(h)$ and $(i)$ 
	are such that the pair of chords is separated by an arbitrary number of $\psi_{I_i}$ and $\Bar{\psi}_{I_i}$ insertions (or H-chord insertions). 
	The pair of chord diagrams in $(a)-(d)$ can share indices,
	hence $p_{ij} \neq 0$. In diagrams $(e)-(g)$, the pair of chords can not share indices i.e. $p_{ij} = 0$,  
        while in diagrams $(h)$ and $(i)$ the chord is such that the
        contracted indices are not shared with any other chord.}
	\label{fig:cSYK_chord_diagram_weights}
    \end{table}
We follow the nomenclature given in \cite{Berkooz:SUSY_SYK} and call the chord diagrams for which $p_{ij}$ can be non-zero, 
``friends'' and those for which $p_{ij}$ has to vanish, ``enemies''. 
Given $k$ insertions of the Hamiltonian, the total number of pairs of chords is
\begin{equation}\label{eq:total_pair_of_chords}
	\#\,\text{ friends} + \#\,\text{ enemies} = \begin{pmatrix}k\\2\end{pmatrix}.
\end{equation}
Additionally, let us restrict to a subspace with a fixed number of left and right pointing chords, $n_\leftarrow$ and $n_\rightarrow$ respectively. 
The number of ways to choose two chords with the same orientation is 
\begin{align}
	\#\,(I.d) + \#\,(I.e) + \#\,(I.g) &= \begin{pmatrix}n_\rightarrow\\2\end{pmatrix},\\
	\#\,(I.b) + \#\,(\overline{I.e}) + \#\,(\overline{I.g}) &= \begin{pmatrix}n_\leftarrow\\2\end{pmatrix}.
\end{align}
Here, a bar over the figure number indicates the opposite orientation compared to \cref{fig:cSYK_chord_diagram_weights}. 
The subspace with fixed $n_\leftarrow,n_\rightarrow$ will be labelled by the sum and difference of $n_\leftarrow,n_\rightarrow$,
\begin{equation}\label{eq:subspace_vars}
	k \equiv n_\leftarrow + n_\rightarrow \;\;\;\text{and} \;\;\; m \equiv n_\rightarrow - n_\leftarrow.
\end{equation}
Using this description, we can now re-express the numbers $\#\,(I.d)$ and $\#\,(I.b)$  as functions of the total number of 
Hamiltonian insertions and the subspace variable $m$, i.e.\
\begin{align}
	\#\,(I.d) &= \begin{pmatrix}\frac{k+m}{2}\\2\end{pmatrix} - \#\,(I.e) - \#\,(I.g),\\
	\#\,(I.b) &= \begin{pmatrix}\frac{k-m}{2}\\2\end{pmatrix} - \#\,(\overline{I.e}) - \#\,(\overline{I.g}),
\end{align}
Thus, \eqref{eq:GC_moments} reduces to
\begin{align}\label{eq:moments_friends_enemies_description} 
	m_k(\mu) &=
	\frac{\mathcal{J}^k}
        {\lambda^\frac{k}{2}}(\cosh\mu)^{N-kp}\sum_{\substack{m
	= -k,-k+2,\\ \cdots,k-2,k}} e^{\mu m}\nonumber\\ 
	&\times\sum_{OCD}\left(\left(\text{diagram weight}\right)
        ^{\lambda \times\#
	\text{friends}}\times\left(\text{diagram weight}\right)^{\lambda\times
        \#\text{enemies}}\right). 
\end{align}
The factors of $(\cosh\mu)^{N-kp}$ and $ e^{\mu m}$ are due to the contributions from chord diagrams $(h)$ and $(i)$. 
From \eqref{eq:total_pair_of_chords}, we get
\begin{equation}\label{eq:friends_enemies_relation}
	\#\text{ enemies} = \begin{pmatrix}k\\2\end{pmatrix} - \#\text{ friends}.
\end{equation}
The contribution from $\#\text{ friends}\times\text{log}(\text{diagram
weight})$ in \eqref{eq:moments_friends_enemies_description} is found
by evaluating 
\begin{eqnarray}
	\lefteqn{\#\text{ friends}\times\text{log}(\text{diagram
 weight}) =}\nonumber \\
 & & \#(I.a)\left( 2 e^{\mu}\cosh\mu -1\right)+\#(I.c)
 \left( 2 e^{-\mu}\cosh\mu -1\right)
 \nonumber\\ & &
 +\left[\begin{pmatrix}\frac{k+m}{2}\\2\end{pmatrix} -
 \#\,(I.e) - \#\,(I.g)\right]\left( 2 e^{-\mu}\cosh\mu -1\right)
 \nonumber \\ & &
	+ \left[\begin{pmatrix}\frac{k-m}{2}\\2\end{pmatrix} -
 \#\,(\overline{I.e}) - \#\,(\overline{I.g})\right]
 \left( 2 e^{\mu}\cosh\mu
 -1\right). \label{eq:moment_friends_contribution}
\end{eqnarray}
The contributions from the second term in (\ref{eq:moments_friends_enemies_description})
is given by $(-1)^{\#\,\text{int}} e^{-\frac{p^2}{N}}$ where $\#\,\text{int}$ counts the number of pairs of chords which intersect.
Using \eqref{eq:friends_enemies_relation} in \eqref{eq:moments_friends_enemies_description} and simplifying further,
we get the $k$th moment expressed as a sum over the subspace of $k$ and $m$ to be
\begin{align}\label{eq:GC_moments_as_subspace_sum}
	m_k(\mu) &= \frac{\mathcal{J}^k}{\lambda^\frac{k}{2}}(\cosh\mu)^{N-kp}\exp\bigg[\lambda\frac{k}{2}(1-\sinh 2\mu)\bigg]\exp\bigg[\lambda\frac{k^2}{2}\sinh^2{\mu}\bigg]\nonumber\\
	&\phantom{=}\sum_{m=-k,\cdots,k} e^{\mu m}\exp\bigg[-\lambda\frac{m}{2}(k-1)\sinh{2\mu}+\lambda\frac{m^2}{2}\cosh^2{\mu}\bigg]m_{k;m}.
\end{align}
Where $m_{k;m}$ represents all the possible chord diagram contributions in the subspace given by $m$ and $k$,
\begin{align}\label{eq:subspace_moment}
	m_{k;m} = &\phantom{=}\sum_{\text{OCD}}(-1)^{\#\,\text{int}}\exp\left(2\lambda e^{-\mu}\cosh\mu[\#\,(I.c)-\#\,(I.e) - \#\,(I.g)]\right)\nonumber\\
	&\phantom{=\sum_{\text{OCD}}(-1)^{\#\,\text{int}}}\times\exp\left(2\lambda e^{\mu}\cosh\mu[\#\,(I.a)-\#\,(\overline{I.e}) - \#\,(\overline{I.g})]\right)
\end{align}
The localized transfer matrix is therefore encoded within $m_{k;m}$.
\subsection{Setting up the Transfer Matrix}\label{subsec:Setting_up_T}
As customary, in the double scaling limit of various SYK like models we will introduce open chord states.
These can be seen as cut-open chord diagrams
and their inner product constitutes a sum over all chord diagrams with the same number of chords \cite{Ambrosini:2024sre}.
In the Majorana SYK, an open chord state is usually denoted as $\ket{n}$, where $n$ is the number of open chords 
(the state does not retain an information on previously closed chords).
However, in our current description chords have an orientation and the order of the sequence in which the differently oriented chords occur is important.
We will therefore have to keep track of that ordering.
A state with zero left and right oriented chords is denoted by $\ket{0}$.
Then, following a similar prescription as given in \cite{Berkooz:SUSY_SYK}, henceforth we use ordered strings of symbols $X$ and $O$, to denote states with open oriented chords.
Each character in the string represents the end of an oriented chord.
\begin{figure}[ht]
    \centering
    
    \begin{tikzpicture}
      \def\n{11}
      \def\spacing{1.15cm} 
    
      \node (s) at (0,0) {$\dots$};
      \draw (0.35,0) -- ({(\n+1)*\spacing +.5},0);
      \node (e) at ({(\n+1.5)*\spacing},0) {$\dots$};
    
      \foreach \i in {1,...,\n} {
        \coordinate (P\i) at ({(\i)*\spacing},0);
      }
    
      \coordinate (A1) at (5*\spacing,1.5*\spacing);
      \coordinate (A3) at (11.5*\spacing,1.5*\spacing);
      \coordinate (A2) at (11*\spacing,.75*\spacing);
      \coordinate (A4) at (11.5*\spacing,.75*\spacing);
     \node[below] at (P1) {$X$};
     \node[below] at (P2) {$O$};
     \node[below] at (P3) {$O$};
     \node[below] at (P4) {$X$};
     \node[below] at (P5) {$O$};
     \node[below] at (P6) {$X$};
     \node[below] at (P7) {$O$};
     \node[below] at (P8) {$X$};
     \node[below] at (P9) {$O$};
     \node[below] at (P10) {$X$};
      \draw[postaction={decorate},
            decoration={markings, mark=at position 0.5 with {\arrow{<}}}]
            (P1) to[bend left=90] (P5);
      \draw[postaction={decorate},
            decoration={markings, mark=at position 0.5 with {\arrow{>}}}](P2) to[bend left=90] (P4);
      \draw[postaction={decorate},
            decoration={markings, mark=at position 0.75 with {\arrow{>}}}] (P3) to[bend left=40] (A1);
            \draw[postaction={decorate},
            decoration={markings, mark=at position 0.5 with {\arrow{>}}}] (A1) to (A3);
       \draw[postaction={decorate},
            decoration={markings, mark=at position 0.5 with {\arrow{<}}}] (P6) to[bend left=90](P7);
      \draw[postaction={decorate},
            decoration={markings, mark=at position 0.5 with {\arrow{<}}}] (P8) to[bend left=90] (P9);
        \draw[postaction={decorate},
            decoration={markings, mark=at position 0.75 with {\arrow{<}}}] (P10) to[bend left=40] (A2);
            \draw[postaction={decorate},
            decoration={markings, mark=at position 0.75 with {\arrow{<}}}] (A2) to (A4);
    
    \end{tikzpicture}
    \caption{Part of a cut-open moment diagram expressed in terms of string of $Xs$ and $O$s}
    \label{fig:XO_diagram}
\end{figure}
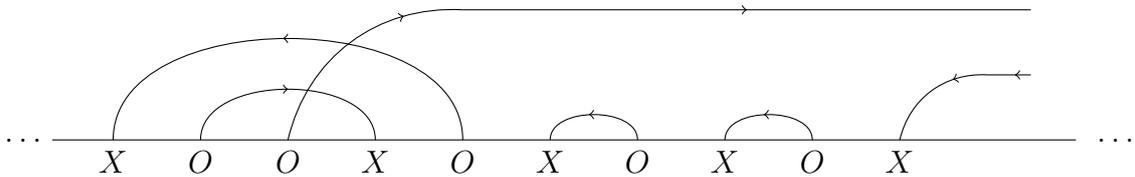

We stipulate that oriented chords emerge from insertion points with label $O$ and end in insertion points with label $X$, 
i.e.\ adding an $O$ to a chord diagram, we can either open a new right pointing chord or close an existing left pointing chord
(the converse holds for adding an $X$). 
An example is given in \cref{fig:XO_diagram}.
Finally, if the state has non-zero $n_X$ and(or) $n_O$ we define $P(n_X,n_O)$ as a string of $n_X \; X$s and $n_O \; O$s of some unspecified but fixed order.
This prescription will be useful in the next section when evaluating inner products.

Using the definitions above, the $k$th moment as a function of the chemical potential is given as
\begin{equation}\label{eq:GC_moment_tansfer_matrix}
	m_{k}(\mu) = \bra{\phi}T^k_\mu\ket{\phi}
\end{equation}
where we tentatively used $T_\mu$, the grand canonical transfer matrix as a function of $\mu$, assuming for now that it is local.
The state $\ket{\phi}$ is its, yet to be defined, ground state. 
It belongs to an infinite dimensional Hilbert space $\mathcal{H}$, called the chord space, with inner product defined in \cref{sec:inner_product}.
The transfer matrix $T_{\mu}$ is defined such that $T_\mu : \mathcal{H}\rightarrow \mathcal{H}$ acts on a given chord basis state, 
$\ket{P(n_X,n_O)}$, by either opening chords, i.e.\ increasing the length of the string, or closing chords with a penalty depending on the number of other open oriented chords intersected by the oriented chord being closed.
The exact form of $T_\mu$ is at this point unknown and will be worked out in \cref{sec:Compare_To_GC}.
As a necessary first step, we shall introduce oriented chord operators that close or open a single chord.

Equation \eqref{eq:GC_moments_as_subspace_sum} describes the grand canonical moment \eqref{eq:GC_moment_tansfer_matrix} as a sum over the possible chord diagram contributions in a given subspace defined by $m$ and $k$ Hamiltonian insertions. 
To ensure locality, we project into a given $m$ subspace, by introducing the auxiliary charge $\theta$.
The expression \eqref{eq:subspace_moment} can then be re-expressed as a Fourier integral
\begin{align}
	m_{k;m} &= \sum_{\text{OCD}}(-1)^{\#\,\text{int}}\exp\big\{2 \lambda e^{-\mu}\cosh\mu[\#\,(I.c) - \#\,(I.e) - \#\,(I.g)]\nonumber\\
	&\phantom{=\sum_{\text{OCD}}(-1)^{\#\,\text{int}}\exp\big\{}+ 2 \lambda e^{\mu}\cosh\mu[\#\,(I.a) - \#\,(\overline{I.e}) - \#\,(\overline{I.g})]\big\}\nonumber\\
	&= \frac{1}{2\pi}\int_0^{2\pi}\dd{\theta} e^{-im\theta}\bra{\phi}T^{k}_\mu(i\theta)\ket{\phi}.
	\label{eq:fixed_m_fourier}
\end{align}
In the Fourier integral, we introduced the transfer matrix on the auxiliary charge subsector, $T_\mu(i\theta)$,
and add additional rules when opening chords \cite{Berkooz:SUSY_SYK}:
\begin{itemize}
	\item when a right pointing chord is opened we multiply the diagram by $ e^{i\theta}$,
	\item when a left pointing chord is opened we multiply the diagram by $ e^{-i\theta}$.
\end{itemize}
Each term in \eqref{eq:fixed_m_fourier} is thus proportional to $e^{in\theta}$ for $n = -k,\cdots,k$.
Therefore, only terms with $m = -n$ contribute, this allows us to extend the sum over $m$ to include all integers.

Next, we introduce the aforementioned oriented chord operators.
First, we define the operator $\alpha$, which adds an insertion point with label $O$, 
i.e.\ acting on a state it either appends an $O$ at the end of the existing string, or it deletes an $X$ from it,
\begin{align}
	\alpha\ket{P(n_X,n_O)} &=  e^{i\theta}\ket{P(n_X,n_O)O} \nonumber\\
	\phantom{=}&+ \sum_{i}W_\mu(P(n_X,n_O)-X_i)\ket{P(n_X,n_O)-X_i} ,
\end{align}
where the sum runs over all positions $i$ with an open $X$ chord. 
In the corresponding state $\ket{P(n_X,n_O)-X_i}$,
the $i$th $X$ chord has been closed.
Put differently, the operator can either open a right pointing chord and multiply the diagram with $ e^{i\theta}$,
or close a left pointing chord with a penalty factor $W_\mu(P(n_X,n_O)-X_i)$.  
Conversely, the operator $\beta$ is defined such that,
\begin{align}
	\beta\ket{P(n_X,n_O)} &= e^{-i\theta}\ket{P(n_X,n_O)X} \nonumber\\
	\phantom{=}&+ \sum_{i}\tilde{W}_\mu(P(n_X,n_O)-O_i)\ket{P(n_X,n_O)-O_i} ,
\end{align}
where now the sum is over positions $i$ with an open $O$ chord. 
In the state $\ket{P(n_X,n_O)-O_i}$, the $i$th $O$ chord has been closed.
The operator $\beta$ can either append an $X$ to the string (open a left pointing chord) and multiply the diagram with $e^{-i\theta}$,
or remove an $O$ from the string (close a right oriented chord) with a penalty factor  $\tilde{W}_\mu(P(n_O,n_X)-O_i)$. 
Finally, we require that
\begin{align}
	\alpha\ket{0} &= e^{i\theta}\ket{O},\nonumber\\
	\beta\ket{0} &= e^{-i\theta}\ket{X},
\end{align}
where the zero-chord state, $\ket{0}$, is the state that is devoid of any chords.
Before proceeding, we shall obtain an expression for $W_\mu(P(n_X,n_O))$ and $\Tilde{W}_\mu(P(n_O,n_X))$.
Both penalty factors for closing an $X$ and an $O$, given by $W_\mu$ and $\tilde{W}_\mu$ respectively, can be obtained 
from \eqref{eq:subspace_moment}.
When closing an $X$ the number $\#\,(I.a)$ is equal to $\# O_{\text{left}}$ (with regard to the $X$ being closed).
While the numbers $\#\,(\overline{I.e})$ and $\#\,(\overline{I.g})$ are equal to $\# X_{\text{left}}$ and $\# X_{\text{right}}$ respectively 
(with regard to the $X$ being closed). 
The number of intersections while closing the $X$ provide the factor of $(-1)^{p^2(\#O_{\text{right}}+\#X_{\text{right}})}$.
Therefore, the penalty factor for closing an $X$ is given by,
\begin{align}\label{eq:expression_for_W}
	W_\mu &= \left(-1\right)^{p^2\left(\#O_\text{right}+\#X_\text{right}\right)}\exp\left[\left(2\lambda 
	e^{-\mu}\cosh\mu\right)\left(\#O_\text{left} - \#X_\text{left} - \#X_\text{right}\right)\right].
\end{align}
Similarly, the penalty factor for closing an $O$ is given by,
\begin{align}\label{eq:expression_for_tilde_W}
	\Tilde{W}_\mu  &= \left(-1\right)^{p^2\left(\#O_\text{right}+\#X_\text{right}\right)}
	\exp\left[\left(2\lambda e^{\mu}\cosh\mu\right)\left(\#X_\text{left} - \#O_\text{left} - \#O_\text{right}\right)\right].
\end{align}
		\subsection{\texorpdfstring{$\boldsymbol{m}\mathbf{=0}$}{TEXT} Sector}
			\label{sec:m0sector}
			In this section, we will restrict our previous considerations to the $m=0$ sector.
This is warranted by the special form of the complex SYK Hamiltonian.
One Hamiltonian insertion 
introduces an ordered pair of a $\psi$ and a $\psi^\dagger$ block each of length $p$.
Furthermore, these pairs also need to be contracted pairwise. 
We will use this fact later for the $H$-chord formalism. 
However, for now we only point out
that this implies an equal number of left and right pointing chords, i.e.\ $m=0$.
As we will see, this fact simplifies our expressions for the moments considerably.

Let us for now stay in the constant auxiliary charge $\theta$ space.
Terms that contribute to the $k$th moment can be represented by closed paths with $2k$ edges
that start and end at the zero chord state $\ket{\Phi}$ in the graph in \cref{fig:physical_hilbert_space}.
We can only change the phase of each term by increments of $\exp(\pm i \theta)$ at a time,
by descending one level in the diagram.
Thus, each term is proportional to a phase $\exp(in\theta)$, where $n=n_\leftarrow - n_\rightarrow$ (counting previously closed chords as well),
referring to the number of the respective creation operators encountered in the path.
After performing the integration in \cref{eq:fixed_m_fourier}, we therefore end up with
\begin{equation}
    m_{k;m} = \sum_{n} \delta_{nm} \text{Paths}(2k,n),
\end{equation}
where $\text{Paths}(2k,n)$ is a sum over all paths with phase $\exp(in\theta)$ and each term is the product of the penalty factors corresponding to a path.
Consequently, only the terms with a vanishing phase,
i.e.\ the paths with equal number of $a^\dagger$ and $b^\dagger$,
contribute to $m=0$ sector.
This is interesting because it implies that once we restrict to $H$-chords our grand canonical transfer matrix is indeed local.
This is not the case for the $\mathcal{N}=2$ SUSY SYK \cite{Berkooz:SUSY_SYK}.

Next, we shall look at the $\mu$-space moments. 
Initially, we introduced these as a sum over all $m$-sectors (see \cref{eq:GC_moments_as_subspace_sum}).
However, we now know that the terms with non-vanishing $m$ do not contribute,
and we can hence rewrite the grand canonical moment as
\begin{align}
    m_k(\mu) &= \frac{\mathcal{J}^k}{\lambda^\frac{k}{2}}(\cosh\mu)^{N-kp}\exp\bigg[\lambda\frac{k}{2}(1-\sinh{2\mu})\bigg]\exp\bigg[\lambda\frac{k^2}{2}\sinh^2{\mu}\bigg] m_{k;0}.
\end{align}
Again, these are now local. 
As in \cite{Berkooz:SUSY_SYK}, we can go to canonical (fixed charge) moments via Fourier transform
\begin{align}
    \label{eq:Q_moment_fourier}
    m_k(Q) &= \frac{1}{2\pi}\int_{-\infty}^{\infty}\dd{\mu}\text{e}^{iQ\mu}m_k(i\mu).
\end{align}
To solve the integral,
we perform a large $N$ saddle point approximation. 
The corresponding calculation is explained in \cref{sec:saddle_appendix}.
Using this,
we get the penalty factors for the canonical transfer matrix with oriented chords $T_{\mu_0}$ in a fixed charge sector (where $\mu_0$ is the saddle point solution given in \eqref{eq:mu_0})
\begin{align}
    W_{Q} &= (-1)^{\#O_\text{right}+\#X_\text{right}}\nonumber\\
	&\phantom{=(}\exp\left[2\lambda\frac{1}{1-2\mathcal{Q}}(  \#O_\text{left} - \#X_\text{left} - \#X_\text{right}  )\right],\\
    \tilde{W}_{Q} &= (-1)^{\#O_\text{right}+\#X_\text{right}}\nonumber\\
	&\phantom{=(}\exp\left[2\lambda\frac{1}{1+2\mathcal{Q}}(  \#X_\text{left} - \#O_\text{left} - \#O_\text{right}  )\right],
\end{align}
where $\mathcal{Q} = Q/N$.
		\subsection{Operator Algebra of the Bosonic Sector}
			\label{sec:bosonic_sector}
We have pointed out before that the structure of the cSYK Hamiltonian requires left- and right-pointing chords be created and closed in pairs.
This implies that states from the two fermionic sectors, namely states of the form $\ket{nX}$ and $\ket{\bar{n}O}$, will never appear
in the calculation for the moments.
To both prepare and ultimately being able to connect the oriented chord formulation to the $H$-chord formulation 
\cite{Berkooz:2020uly},
we introduce bosonic operators by combining fermionic ones.

To this end, we shall first split our operators $\alpha$ and $\beta$ into creation and annihilation operators for $X$s and $O$s, i.e.
\begin{align}
    \alpha &\equiv a^\dagger + b, \\
    \beta &\equiv a + b^\dagger,
\end{align}
where the action of $a^\dagger$($b^\dagger$) is to insert an $O(X)$ at the end of the string 
and the action of $a(b)$ is to delete an $O(X)$ from the start or end of the string. 
The reader can verify that $a^\dagger$ is indeed the adjoint of $a$ and the same holds for $b$. 
The combined operators $ab$ and $ba$ act on the bosonic sector of the physical Hilbert space with $n>1$ as follows:
\begin{align}
        ab\ket{n} &= 
        \begin{aligned}[t] &-\ket{(n-1)}\exp\left[
            -4\lambda(n-1)\cosh^2\mu
        \right]\\
        &-\ket{\overline{(n-1)}} \exp\left[
            -2\lambda(n-1)e^{-\mu}\cosh\mu
        \right],
    \end{aligned}\\
    ab\ket{\overline{n}} &= 
        \begin{aligned}[t]
        &+\ket{(n-1)}\exp\left[
            2\lambda \left(e^{-\mu} -[n-1]e^\mu  \right)\cosh\mu
        \right]
        \\ &+\ket{\overline{(n-1)}}\exp\left[
            2\lambda e^{-\mu}\cosh\mu
        \right],
        \end{aligned}\\
        ba\ket{n} &= 
        \begin{aligned}[t] &+\ket{(n-1)}\exp\left[
            2\lambda e^{\mu}\cosh\mu
        \right]\\
        &+\ket{\overline{(n-1)}} \exp\left[
            2\lambda \left(e^{\mu} -[n-1]e^{-\mu}  \right)\cosh\mu
        \right],
    \end{aligned}\\
    ba\ket{\overline{n}} &= 
        \begin{aligned}[t]
        &-\ket{\overline{(n-1)}}\exp\left[
            -4\lambda(n-1)\cosh^2\mu
        \right]
        \\ &-\ket{(n-1)}\exp\left[
            -2\lambda(n-1)e^{\mu}\cosh\mu
        \right].
        \end{aligned}
\end{align}
The case $n=1$ is special, we have 
\begin{align}
    ab \ket{1} = - \ket{0},\\
    ba \ket{1} = \exp\left[2\lambda e^\mu \cosh(\mu)\right]\ket{0}
\end{align}
and similar relations for $\ket{\overline{1}}$.

Next, we introduce the bosonic number operator which simply counts the number of $XO(OX)$ pairs,
\begin{align}
    \hat{N}\ket{n} = n\ket{n},\quad \hat{N}\ket{\overline{n}} = n\ket{\overline{n}},
\end{align}
and the two bosonic annihilation operators
\begin{align}
    A\equiv ab + q_1 ba,\quad B\equiv ba + q_2 ab,\label{eq:bosonic_ops}
\end{align}
with $q_1(q_2)$ to be determined such that $A(B)$ is an endomorphism on the sector $\{\ket{n}\}$
$(\{\ket{\overline{n}}\})$, in the sense that
\begin{align}
    A\ket{n} = c_1(\hat{N})\ket{(n-1)},\\
    B\ket{\overline{n}} = c_2(\hat{N})\ket{\overline{(n-1)}}.
\end{align}
By equating coefficients, each of the above relations yields two equations,
which we solve to find
\begin{align}
    q_1 &= \exp\left[-2\lambda\cosh\mu e^\mu\right],\\
    q_2 &= \exp\left[-2\lambda\cosh\mu e^{-\mu}\right],
\end{align}
and
\begin{align}
    c_1(N) = c_2(N) &= 1-\exp\left[ -4\lambda \cosh^2\mu \hat{N}\right],\\
    &= (1-q(\mu))\left[\hat{N}\right]_{q(\mu)}
\end{align}
where $q(\mu)\equiv \exp(-4\lambda\cosh^2\mu)$ 
and as usual $[n]_q\equiv(1-q^n)(1-q)^{-1}$ is the $q$-deformed integer (see e.g.\ \cite{Lin:2023trc}).
For $n\ge1$, we also introduce normalized versions of $A$ and $B$:
\begin{align}
  \tilde{A}&\equiv \left(
        1-q\left(\mu\right)^{\hat{N}}
    \right)^{-1}\left(\exp\left[-2\lambda\cosh\mu e^\mu\right]ba + ab\right),\\
   \tilde{B}&\equiv \left(
         1-q\left(\mu\right)^{\hat{N}}
    \right)^{-1}\left(\exp\left[-2\lambda\cosh\mu e^{-\mu}\right]ab + ba\right).  
\end{align}
Using this, it becomes immediately clear that 
\begin{align}
    \tilde{A}a^\dagger b^\dagger\ket{n}&=\ket{n},\\
    \tilde{B}b^\dagger a^\dagger\ket{\overline{n}}&=\ket{\overline{n}}.
\end{align}
But as the action of $a^\dagger b^\dagger$ and $b^\dagger a^\dagger$ is to append a pair $XO$ or $OX$ respectively,
we can conclude that $\tilde{A}$ closes a pair $XO$ when acting on $\ket{n}$ and $\tilde{B}$ closes a pair $OX$ when acting on $\ket{\overline{n}}$.
Moreover, notice $a^\dagger b^\dagger$ and $b^\dagger a^\dagger$ send all vectors with a string of pairs of the opposite order to the null vector,
\begin{align}
    \ker\left(a^\dagger b^\dagger\right)&=\vecspan\left(\{\ket{\overline{n\ge 1}}\}\right),\\
    \ker\left( b^\dagger a^\dagger\right)&=\vecspan\left(\{\ket{n\ge 1}\}\right).
\end{align}
In contrast, acting with $A^\dagger$/$B^\dagger$ on $\ket{0}$ produces a combination of $\ket{1}$ and $\ket{\overline{1}}$.
Additionally, notice that $A\ket{1}=B\ket{\overline{1}} =0$.
Thus, the ground state of the bosonic sectors is $\ket{1}$/$\ket{\overline{1}}$ respectively, while the zero chord state of oriented chords, $\ket{0}$, is therefore not part of the bosonic sector.
In consequence, we set $\ket{\phi}\equiv \ket{1}$, without loss of generality.
The bosonic states $\vecspan\{\ket{n\ge 1}\}$/$\vecspan\{\ket{\overline{n\ge 1}}\}$ are then closed under the action of $A^{(\dagger)}$ and $B^{(\dagger)}$ respectively.
We have now managed to decouple the two bosonic sectors.

We shall now focus on the space spanned by $\{\ket{n\ge 1}\}$ 
and write down the algebra generated by acting on it with the operators $N$, $A^\dagger \equiv q_1 a^\dagger b^\dagger + b^\dagger a^\dagger $  
and $A$:
\begin{align}
    \left[A,A^\dagger \right]_{q(\mu)}&=q_1(1-q(\mu)),\label{eq:oriented_algebra1}\\
    \left[N,A^\dagger \right]&=A^\dagger, \label{eq:oriented_algebra2}\\
    \left[N, A \right] &= -A.\label{eq:oriented_algebra3}
\end{align} 
We will later use these relations to compare to the algebra of the $H$-chord formalism.
To that end, we also give the commutators in a charge subsector with normalized charge $\mathcal{Q}$. 
The only thing we need to modify is replacing $q_1$ by $\exp(-2\lambda\frac{1}{1+2\mathcal{Q}})$ everywhere and the first equation becomes
\begin{align}
    \left[A_Q,A_Q^\dagger \right]_{q(Q)}&=q_1(1-q(Q)),
\end{align}
where $q(Q)\equiv\exp\left(-4\lambda\frac{1}{1-4\mathcal{Q}^2}\right)$.
At this stage, we can normalize the operators such that $[\tilde{A}_Q,\tilde{A}^\dagger_Q] = 1$ i.e.\ $\tilde{A}_Q = \frac{A_Q}{\sqrt{q_1(1-q(Q))}}$ and $\tilde{A}^\dagger_Q = \frac{A^\dagger_Q}{\sqrt{q_1(1-q(Q))}}$.
Acting with $\tilde{A}_\mathcal{Q}$ and $\tilde{A}^\dagger_\mathcal{Q}$ on the normalized state $\ket{\tilde{n}}$ (as defined in \eqref{eq:norm_states}) we get that,
\begin{align}
    \tilde{A}_Q\ket{\widetilde{n}} &= \frac{c_1(n)}{\sqrt{A_{n+1}}\sqrt{q_1(1-q(Q))}}\ket{n} = \frac{c_1(n)}{\sqrt{q_1(1-q(Q))}}\sqrt{\frac{A_{n}}{A_{n+1}}}\ket{\widetilde{n-1}},\nonumber\\
    &= \sqrt{[n]_{q(Q)}}\ket{\widetilde{n-1}},\\
    \tilde{A}^\dagger_Q\ket{\tilde{n}} &= \frac{q_1}{\sqrt{A_{n+1}}\sqrt{q_1(1-q(Q))}}\ket{n+2} = \sqrt{\frac{q_1}{1-q(Q)}}\sqrt{\frac{A_{n+2}}{A_{n+1}}}\ket{\widetilde{n+1}},\nonumber\\
    &= \sqrt{[n+1]_{q(Q)}}\ket{\widetilde{n+1}}.
\end{align}
Where for $n \ge 1$ $A_{n}$, can be found via the recursion relation \eqref{eq:recursion_An} and $A_0=1$.
Notice that we also performed a shift in $n$, i.e.\ $\tilde{n}$ ranges from $0$ to $\infty$.
From here on, we shall always use the shifted version and drop the tilde.
	\section{Transfer Matrix and Krylov Complexity}
		\label{sec:T_Matrix_K_Comp}
		In this section, we compute the Krylov complexity in the grand canonical and the canonical ensembles.
In \cref{sec:fixedQTransferMatrix}, we obtain the transfer matrix associated to the $H$-chords.
Furthermore, we define the algebra for the $H$-chord creation and annihilation operators as well as 
their commutation relations with the number operator, charge creation and annihilation operators.
In \cref{sec:Compare_To_GC}, we write down the relation between the $k^{\text{th}}$ moment in the grand canonical 
ensemble and the $H$-chord transfer matrix.
Moreover, we show that the algebra of the $H$-chord operators matches that of the oriented chord operators and thus 
the $H$-chords can be used to express the grand canonical transfer matrix.
Finally, we provide expressions for the Krylov complexities in the grand canonical and canonical ensembles. 
The grand canonical Krylov complexity can be expanded as a weighted sum over the canonical
Krylov complexities.   
		\subsection{Transfer Matrix in Fixed Charge Sector}
			\label{sec:fixedQTransferMatrix}
In \Cref{sec:saddle_appendix}, we perform a Fourier transform of $m_k(\mu)$ to find the $k^{\text{th}}$ moment in the 
canonical (fixed charge) ensemble.
The result is given in \cref{eq:m_of_Q}.
As explained in \cref{sec:m0sector}, only the terms with $m=0$ contribute and thus the expression further reduces to
\begin{align}
    m_k(Q)
    &\approx \frac{\mathcal{J}^k}{\lambda^\frac{k}{2}} \exp\left[
        N\left(\!-\frac{1}{2} \log \left(\frac{1-4\mathcal{Q}^2}{4}\right)-2\mathcal{Q} \artanh(2\mathcal{Q})\right)
        +\frac{1}{2} k \sqrt{\lambda N } \log \left(1-4\mathcal{Q}^2\right)
   \right]\nonumber\\
    &\hphantom{=}\times\sqrt{\frac{1}{8\pi N\left(1-4\mathcal{Q}^2\right)}}\exp\left[
        -\frac{k \lambda  (4\mathcal{Q} (\mathcal{Q}+1)-1)}{2 \left(1-4\mathcal{Q}^2\right)}
        \right] m_{k;0}(\mu_0/2).
\label{eq:fixed_Q_moment_final_exp}
    \end{align}
By comparing \eqref{eq:fixed_Q_moment_final_exp} to the expression for the fixed charge moment as obtained in \cite{Berkooz:2020uly}, 
we see that $\bra{0}T_H^k(\mathcal{Q})\ket{0}$ in \eqref{eq:fixed_Q_moment_final_exp} 
describes the integral over $\theta$, i.e.\
\begin{align}\label{eq:T_H_matrix_element}
    \bra{0}T_H^k(\mathcal{Q})\ket{0} = &\int_{0}^{\pi}\frac{\dd{\theta}}{2\pi}(q(\mathcal{Q}),\text{e}^{\pm 2i\theta};q(\mathcal{Q}))_{\infty}\left[\frac{2\mathcal{J}\cos(\theta)}{\sqrt{\lambda(1-q(\mathcal{Q}))}}\right]^k\nonumber \\
    &\times \exp\left[\frac{1}{2} k \sqrt{\lambda N } \log \left(1-4\mathcal{Q}^2\right) - \frac{k \lambda  (4\mathcal{Q} (\mathcal{Q}+1)-1)}{2 \left(1-4\mathcal{Q}^2\right)}\right].
\end{align}
In \cite{Berkooz:2020uly}, it was pointed out that therefore the model must have the same $q$-Gaussian density of states as in the Majorana SYK model although in each charge subsector individually.
In addition, the exponential term in the second line of \eqref{eq:T_H_matrix_element} simply rescales the energy.
Thus, in the $H$-chord formalism, the transfer matrix within a fixed charge sector of the cSYK model is 
\begin{align}
    T_H(\mathcal{Q}) = &\frac{\mathcal{J}}{\sqrt{\lambda}}\exp\left\{\left[\frac{\sqrt{\lambda N}}{2}\log\left(1-\frac{4\mathcal{Q}^2}{N^2}\right) 
    + \frac{\lambda}{2}\left(1-\frac{4\mathcal{Q}N}{N^2-4\mathcal{Q}^2}\right)\right]\right\}\nonumber\\
    &\times\left(h_\mathcal{Q}^\dagger 
    + W_l(\mathcal{Q})\eta_\mathcal{Q}\right), \label{eq:H_Chord_Transfer_Matrix}
\end{align}
where, $h_\mathcal{Q}^\dagger$ is the $H$-chord creation operator and $\eta_\mathcal{Q}$ is the $H$-chord annihilation operator, 
while $W_l(\mathcal{Q})$ describes the penalty factor for closing $H$-chords.
These operators act on a state in a given charge subsector such that,
\begin{align}
    h_\mathcal{Q}^\dagger\ket{n} &= \ket{n+1},\nonumber\\
    \eta_\mathcal{Q}\ket{n} &= W_n(\mathcal{Q})\ket{n-1},\;\;\;\; \eta_\mathcal{Q}\ket{0} = 0.
\end{align}
Here, states $\ket{n}$ belong to the auxiliary Hilbert space of chords in a charge subsector $\mathcal{H}_Q = \vecspan{\{\vec{0},\ket{i}\}}_{i\in \mathbb{N}}$.
Furthermore, by promoting the charge $Q$ to an operator it can be readily seen that any state $\ket{n}\in \mathcal{H}_Q$ is an eigenstate of the charge.

Following the procedure described in \cite{Lin:2022rbf} for the Majorana SYK model, the transfer matrix in \eqref{eq:H_Chord_Transfer_Matrix} can be made symmetric to obtain,
\begin{align}\label{eq:Canonical_T_matrix}
    T_H(\mathcal{Q}) &= \frac{\mathcal{J}}{\sqrt{\lambda}}\exp\left\{\left[\frac{\sqrt{\lambda N}}{2}\log\left(1-\frac{4\mathcal{Q}^2}{N^2}\right) + \frac{\lambda}{2}\left(1-\frac{4\mathcal{Q}N}{N^2-4\mathcal{Q}^2}\right)\right]\right\}\nonumber\\
    \phantom{=\exp}&\times\left(\sqrt{\frac{1-q(\mathcal{Q})^{\hat{l}}}{1-q(\mathcal{Q})}}h^\dagger_\mathcal{Q} + h_\mathcal{Q}\sqrt{\frac{1-q(\mathcal{Q})^{\hat{l}}}{1-q(\mathcal{Q})}}\right),
\end{align}
where now the action of the operators and their commutation relation in a given charge subsector is such that
\begin{align}
    h^\dagger_\mathcal{Q}\ket{n} &= \sqrt{[n+1]_{q(\mathcal{Q})}}\ket{n+1},\nonumber\\
    h_\mathcal{Q}\ket{n} &= \sqrt{[n]_{q(\mathcal{Q})}}\ket{n},\\
    \left[h_\mathcal{Q},h^{\dagger}_\mathcal{Q}\right]_{q(\mathcal{Q})} &= 1.\nonumber
\end{align}
We further define a chord number operator where,
\begin{equation}\label{eq:num_op_Can}
    \hat{n} = \sum_{n}n\ketbra{n}
\end{equation}
is defined for a given charge sector and has eigenvalue $n$ when acting on $\ket{n}$ in said charge subsector.
Moreover, the $H$-chord creation (annihilation) and $H$-chord number operator satisfy the following commutation relations:
\begin{align}\label{eq:H_chord_algebra}
	\left[\hat{n},h^{\dagger}_\mathcal{Q}\right] &= h^{\dagger}_\mathcal{Q}, \\
	\left[\hat{n},h_\mathcal{Q}\right] &= -h_\mathcal{Q}.
\end{align}

		\subsection{Comparison to the Grand Canonical Ensemble}
			\label{sec:Compare_To_GC}
Now that we have an expression for the canonical transfer matrix in the H-chord formalism, 
we can infer the structure of the grand canonical transfer matrix $T_\mu$ expressed via oriented chord operators.
We define the grand canonical transfer matrix in analogy with the canonical transfer matrix, 
i.e.\ we can use it to re-express the ensemble average of the grand canonical partition function in terms of a transition amplitude \cite{Rabinovici:2023yex}
\begin{equation}
	Z=\braket{\Tr\left( e^{-\beta H + \mu Q}\right)}_J = \bra{0} e^{-\beta T_\mu} \ket{0},
\end{equation}
where  $\ket{0}$ zero denotes the zero chord state. 
Likewise, we can use the canonical transfer matrix 
derived via the $H$-chord formalism 
in \cref{sec:fixedQTransferMatrix} 
to re-express the canonical partition function
\begin{equation}
	Z_Q=\braket{\Tr_Q\left( e^{-\beta H }\right)}_J  = \bra{0} e^{-\beta T_H} \ket{0},
\end{equation}
where the subscript at the trace indicates that it is taken only 
over states of charge  $Q$ and $T_H= h^\dagger_Q + h_Q$ is the 
transfer matrix in charge subsector $Q$ 
introduced in \cref{sec:fixedQTransferMatrix} 
using $H$-chord operators. 
For convenience, we will now shift all the dependence on the 
parameters $Q$ and $\mu$ to the states.
As detailed in appendix \ref{sec:parameters_appendix} this is
done by assigning a charge eigenvalue $Q$ to a state and 
replacing the $Q$ dependence in operators by a dependence on 
an operator $\hat{Q}$. 
The action of an operator on a state $\left| n, Q\right>$ depends on its eigenvalue $Q$.
For operators, we suppress the $\hat{Q}$ dependence and
the subscript $\mu$ in the notation. 
The grand partition function is just the sum over the canonical partition sums with $\mu$ dependent weights.
Therefore, in our new picture we have
\begin{equation}
	\ket{0, \mu} = \int_{-1/2}^{1/2} \dd{\mathcal{Q}}e^{-\beta \frac{N}{2} \Omega} \ket{0,\mathcal{Q}},
\label{eq:fourstat}
\end{equation}
where $\Omega$ is the grand potential of the complex SYK model in the large $p$ limit \cite{Davison:2016ngz} 
(again, an explicit derivation of this can be found in \cref{sec:parameters_appendix}).
On the level of moments, this yields the following equation
\begin{align}
	m_{2k}(\mu) &= \bra{0,\mu} T^{2k} \ket{0,\mu} \nonumber\\
	&= \int_{-1/2}^{1/2} \dd{\mathcal{Q}}\dd{\mathcal{\tilde{Q}}}e^{-\beta N \Omega}\bra{0,Q} T^{2k}\ket{0,\tilde{Q}} \nonumber\\
	&\stackrel{!}{=} \int_{-1/2}^{1/2} \dd{\mathcal{Q}} m_k(Q) e^{-\beta N \Omega}\\
	&= \int_{-1/2}^{1/2} \dd{\mathcal{Q}} \bra{0,Q} T_H^k e^{-\beta N \Omega} \ket{0,\tilde{Q}}\nonumber\\
	&= \bra{0,\mu} T_H^k \ket{0,\mu}\nonumber
\end{align}
(keep in mind that one $H$-chord is constituted by two oriented chords). 
But as $T$ is block diagonal, it follows that
\begin{align}\label{eq:GC_T_and_C_T_relation}
	\bra{0,\mathcal{Q}} T^{2k}  \ket{0,\tilde{\mathcal{Q}}}=\bra{0,\mathcal{Q}} T_H^k  \ket{0,\tilde{\mathcal{Q}}}.
\end{align}
To reach complete equality between $T^{2k}$ and $T_H^k$, also their action on states with higher chord number must match.
To confirm this, we consider the algebra of $h$ and 
$h^\dagger$ and then identify combinations of oriented 
chord operators with the same commutator algebra.

At $N\rightarrow\infty$, 
the grand canonical ensemble becomes identical to the 
Majorana SYK with re-scaled energy levels \cite{Berkooz:2020uly}, 
which is also reflected by our findings in 
\cref{sec:fixedQTransferMatrix}. 
We would like to find an equivalent version for the grand 
canonical case as given in \cref{sec:fixedQTransferMatrix} for 
the auxiliary Hilbert space in a charge subsector. 
To do so, we note that the $\mathcal{H}_Q$s are separable and 
infinite dimensional (at least for $-\infty < Q < \infty$,) and 
the chord basis is orthonormal \cite{Rabinovici:2023yex,Ambrosini:2024sre},
i.e.\ we can define a unitary isomorphism $c_j: \mathcal{H}_Q \rightarrow \mathcal{H}_{Q+j}$ s.t. $c_j \ket{n,Q}= \ket{n,Q+j}$
and its inverse is given by $c_j^{-1}=c_j^{\dagger}\equiv c_{-j}$. 
Next, we notice that
\begin{align}
	\left[c_j,h^{\dagger}\right]\ket{n,Q} &= \left(\sqrt{[n+1]_{q(Q)}}-\sqrt{[n+1]_{q(Q+j)}}\right)\ket{n+1,Q+j} \nonumber\\
	&=c_j h^{\dagger}\left(1-\frac{\sqrt{[n+1]_{q(Q+j)}}}{\sqrt{[n+1]_{q(Q)}}}\right)\ket{n,Q}
\end{align}
and analogously
\begin{equation}
	\left[c_j,h\right]\ket{n,Q} = c_j h\left(1-\frac{\sqrt{[n]_{q(Q+j)}}}{\sqrt{[n]_{q(Q)}}}\right)\ket{n,Q}.
\end{equation}
In total, we get the extended algebra
\begin{align}
	\left[h,h^{\dagger}\right]_q &= 1,\\
	\left[\hat{n}_Q,h^{\dagger}\right] &= h^{\dagger}, \\
	\left[\hat{n}_Q,h\right] &= -h, \\
	\left[c_j,h^{\dagger}\right] &= c_j h^{\dagger}\left(1-\frac{\sqrt{[n+1]_{q(\hat{Q}+j)}}}{\sqrt{[n+1]_{q(\hat{Q})}}}\right),\\
	\left[c_j,h\right] &=  c_j h\left(1-\frac{\sqrt{[n]_{q(\hat{Q}+j)}}}{\sqrt{[n]_{q(\hat{Q})}}}\right), \\
	\left[c_j,\hat{Q}\right] &= -jc_j,
\end{align}
all other commutators  vanish. 
We can build a representation of this algebra by acting with the chord creation operator and the charge shift operator $c_j$ on the state $\ket{0}$, 
i.e.\ our module is spanned by $\{\vec{0},\Ket{n,Q}\}_{n\in\mathbb{N}}^{Q \in [-N/2, N/2]}$, where $\vec{0}$ is the null vector.
An inner product on this space is defined naturally via the inner product on the charge subspaces, 
if we demand $\braket{n,Q|\tilde{n},\tilde{Q}} = \delta_{n\tilde{n}} \delta_{Q \tilde{Q}}$.
The resulting space is the Hilbert space of the grand canonical ensemble $\mathcal{F}=\bigoplus_{Q\in [-N/2, N/2]} \mathcal{H}_Q$.
Using the basis defined above, we can also define a grand canonical number operator
\begin{align}\label{eq:num_op_GC}
	\hat{n} \equiv \sum_n \sum_Q n\ketbra{n, Q}.
\end{align}
It is easy to verify, that it fulfils the same commutation relation with $h^{(\dagger)}$ as $n_Q$, i.e.
\begin{align}
	\left[ \hat{n}, h^\dagger \right] &= h^{\dagger},\\
	\left[\hat{n},h\right] &= -h.
\end{align}
Now, we can map our states to the oriented chord states defined in \cref{sec:bosonic_sector}. 
The commutator algebra found above matches exactly the algebra for $\tilde{A}$ and $\tilde{A}^\dagger$, \crefrange{eq:oriented_algebra1}{eq:oriented_algebra3}.
Consequently,
\begin{equation}
	T^{2} = \tilde{A} + \tilde{A}^\dagger
\end{equation}
and grand canonical states of chord number $n>0$ can be written as
\begin{align}
	\ket{n,\mu} = \frac{1}{\sqrt{\mathcal{N}}}\int_{-1/2}^{1/2}\dd{\mathcal{Q}} e^{-\beta \frac{N}{2} \Omega} \ket{n,\mathcal{Q}},
\end{align}
with the normalization factor $\mathcal{N}$ given in \cref{eq:normalization}.
		\subsection{Krylov Complexity}
			\label{sec:Krylov}
Having obtained a relation between the states and the transfer matrices in the grand canonical and the canonical ensembles, 
we can now calculate the Krylov complexity in both ensembles and check their relation \eqref{eq:krylov_inequality} as stated in \cite{Caputa:2025mii}.
\begin{figure}[hbtp]
    \centering
    \includegraphics[width=0.7\textwidth]{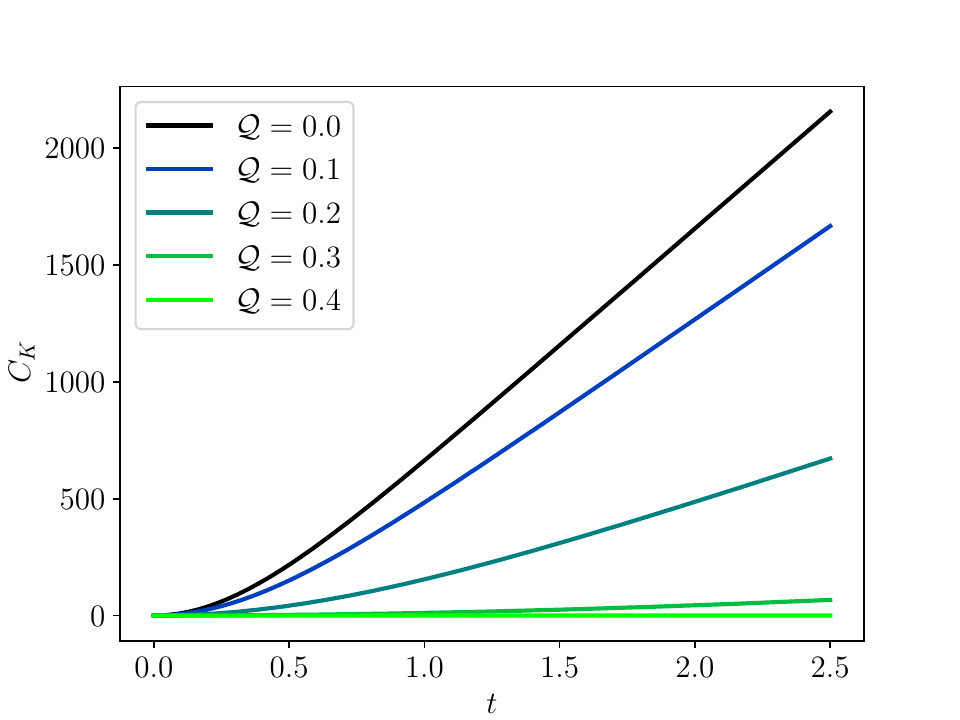}
    \caption{Numerical results for $C_{\mathcal{K}}^{(\mathcal{Q})}(t)$ for different values of $\mathcal{Q}$ at $p = 10$, $\lambda=0.001$ and $\mathcal{J}=1$.}
    \label{fig:C_K_multiQ}
\end{figure}
\begin{figure}[hbtp]
    \centering
    \includegraphics[width=0.7\textwidth]{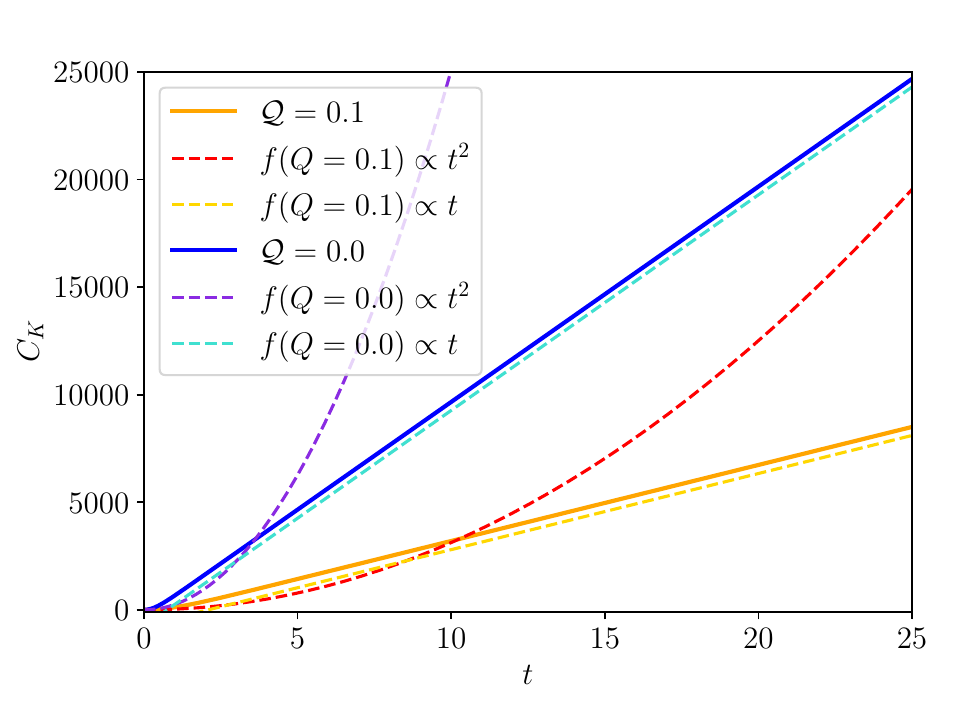}
    \includegraphics[width=0.7\textwidth]{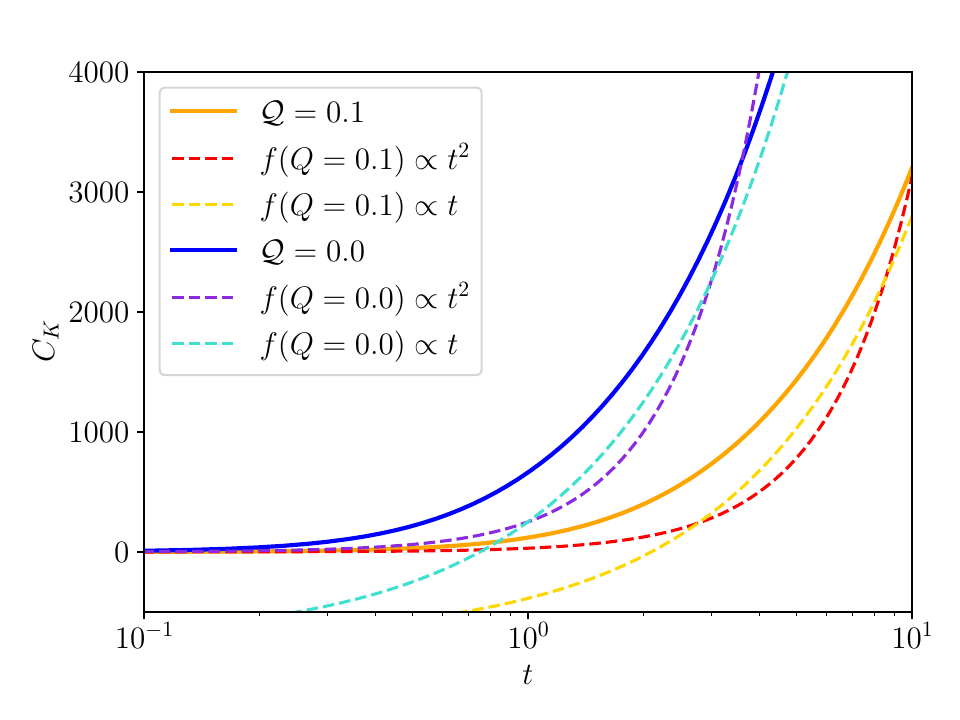}
    \caption{We compare the linear and quadratic asymptotes (dashed lines) for $C_{\mathcal{K}}^{(\mathcal{Q})}(t)$ given in \cref{eq:kryltwo} to the numerical results (solid lines).
    The parameters for the plot are $p = 50$, $\lambda=0.001$ and $\mathcal{J}=1$. 
    The two figures show the same data. 
    However, the bottom plot has a log scale on the $x$-axis.
    }
    \label{fig:C_K_Q_analytic_vs_numerical}
\end{figure}
\begin{figure}[hbtp]
    \centering
    \includegraphics[width=0.7\textwidth]{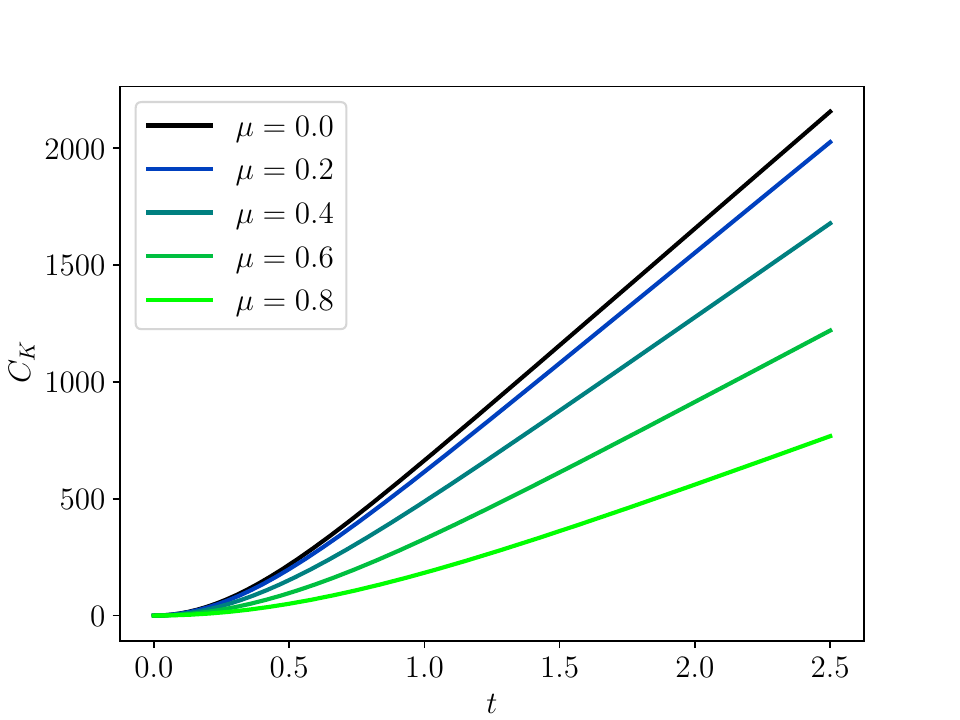}
    \caption{Numerical results for $C_{\mathcal{K}}(t)$ calculated via the $\mu$ dependent Lanczos coefficients \eqref{eq:Lanczos_coeff_GC} for different values of $\mu$ at $p = 10$, $\lambda=0.001$ and $\mathcal{J}=1$.}
    \label{fig:C_K_multi_mu}
\end{figure}
\begin{figure}[hbtp]
    \centering
    \includegraphics[width=0.7\textwidth]{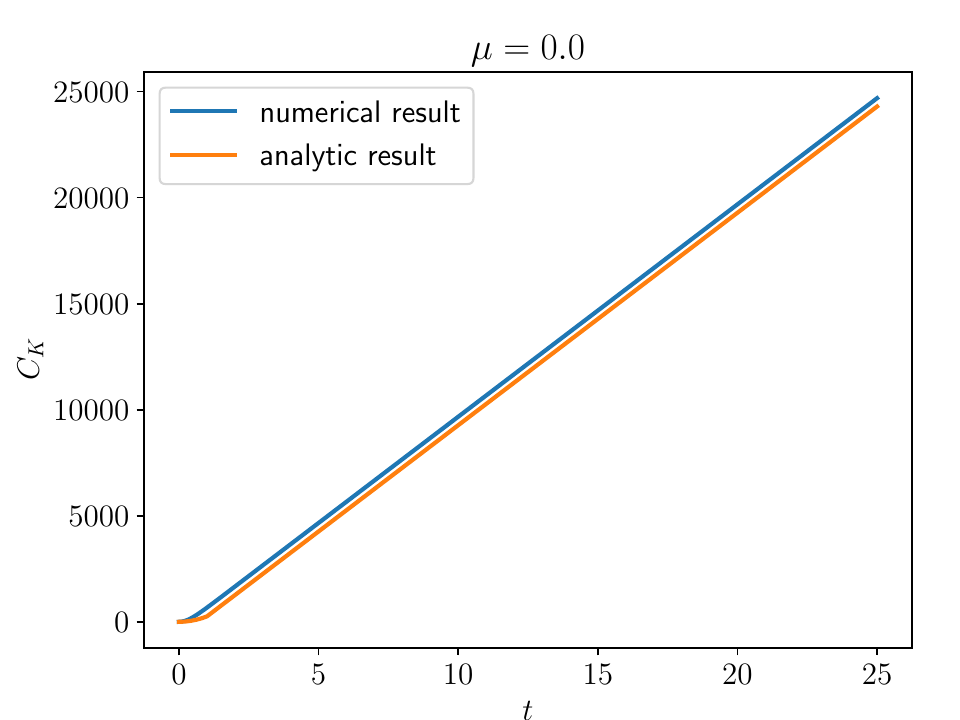}
    \includegraphics[width=0.7\textwidth]{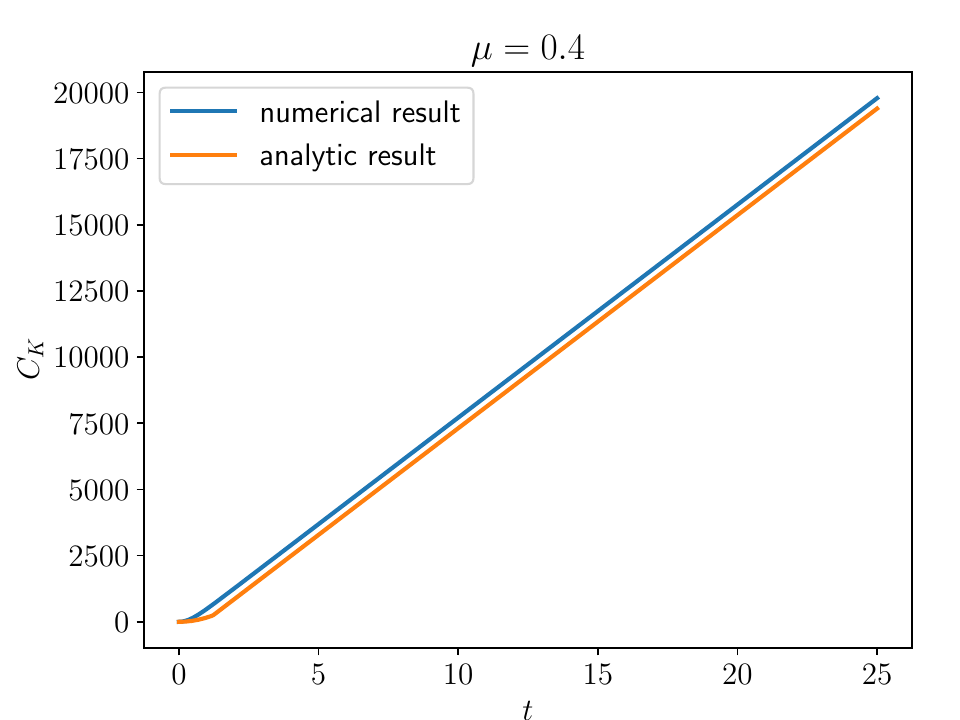}
    \caption{We compare the numerical results for $C_{\mathcal{K}}(t)$ to analytic results obtained by integrating \cref{eq:kryltwo} weighted with $\exp\left[-N\beta\Omega\right]$ as explained in \cref{sec:krylov_in_gc}.}
    \label{fig:C_K_mu_analytic_vs_numerical}
\end{figure}

\subsubsection{Krylov Complexity in the Grand Canonical Ensemble}\label{sec:krylov_in_gc}
A state $\ket{\Psi(\mu,t)}$ in the grand canonical ensemble at arbitrary time $t$ can be defined as,
\begin{equation}
    \ket{\Psi(\mu,t)} = \text{e}^{-itT}\ket{0,\mu}.
\end{equation}
The number operator in the grand canonical ensemble is given by \eqref{eq:num_op_GC}.
Using this, the grand canonical state Krylov complexity is defined as:
\begin{align}\label{eq:GC_Can_Krylov_relation}
    C_\mathcal{K} &= \bra{0}\text{e}^{itT}\hat{n}\text{e}^{-itT}\ket{0} \nonumber\\
    &= \frac{1}{\mathcal{N}} \sum_n\int_{-1/2}^{1/2} \dd{\mathcal{Q}}\dd{\tilde{\mathcal{Q}}}\dd{\mathcal{Q}'}n \text{e}^{-N\beta\Omega(\mu)}\bra{0,\mathcal{Q}}\text{e}^{itT}\ket{n,\tilde{\mathcal{Q}}}\bra{n,\tilde{\mathcal{Q}}}\text{e}^{-itT}\ket{0,\mathcal{Q}'}\nonumber\\
    &= \frac{1}{\mathcal{N}} \int_{-1/2}^{1/2}\dd{\mathcal{Q}} \text{e}^{-N\beta\Omega(\mu)}\bra{0,Q}\text{e}^{itT}\hat{n}_\mathcal{Q}\text{e}^{-itT}\ket{0,\mathcal{Q}} = \frac{1}{\mathcal{N}} \int_{-1/2}^{1/2}\dd{\mathcal{Q}}\text{e}^{-N\beta\Omega(\mu)} C_{\mathcal{K}}^{\mathcal{Q}}.
\end{align}
Where we have used the fact that the transfer matrix in the grand canonical ensemble is block diagonal.
Therefore, from \eqref{eq:GC_Can_Krylov_relation} we can see that the Krylov complexity in the grand canonical ensemble can be obtained as a sum over the Krylov complexities for the individual charge subsectors.
Note, that \eqref{eq:GC_Can_Krylov_relation} does not hold universally.
It depends crucially on the choice of the seed state \eqref{eq:fourstat} and the transfer matrix preserving charge. 
Charge conservation in the underlying SYK model only demands $\bra{0, Q} T^m \ket{0, Q^\prime} \sim \delta_{QQ^\prime}$. 
In \cref{app:charge_mixing}, we comment on this for the case of a simplified model akin to the set-up that resulted from fixing $m$ in \cref{subsec:Setting_up_T}.
\Cref{eq:GC_Can_Krylov_relation} would not hold if $\bra{n,Q} T^m\ket{0,Q^\prime}$ is non-vanishing for $Q \not= Q^\prime$ for any chord number $n$. 
We expect this to happen for the $\mathcal{N}=2$ SUSY SYK.
To associate charge $Q$ with a symmetry transformation we notice that the saddle point equation relates $Q$ to $m$, 
the difference in the number of open chords of opposite orientation. 
This means that for fixed saddle point value of the chemical potential, 
$Q$ can change only if $m$ changes.
On the other hand, the number $m$ is the charge under constant shifts in the auxiliary quantity $\theta$ which can therefore be viewed as the associated symmetry transformation.
In the double-scaled cSYK, charge is manifestly conserved within the $H$ chord formalism. 
For the ${\cal N} =2$ supersymmetric SYK model each insertion of the Hamiltonian in the trace also corresponds to a pair of chords with opposite orientation as long as they are both opened or closed. 
But it could also happen that one chord is closed whereas the other opens. 
In this case, both chords have the same orientation and $Q$ changes during the transfer corresponding to that (double) node. 
However, for the supersymmetric SYK no local transfer matrix is available in the grand canonical setting. 
Therefore, strictly speaking, our arguments do not really apply and further studies are needed.

In the next section, we will verify the relation \eqref{eq:GC_Can_Krylov_relation} numerically.
For this, we will need the Lanczos coefficients as functions of the chemical potential, which are given as
\begin{equation}\label{eq:Lanczos_coeff_GC}
    b_n\left(\tilde{\mu}\right) = \frac{\mathcal{J}}{\sqrt{\lambda}\cosh^p(\tilde{\mu})}\exp\left[\frac{\lambda}{2}(1-\sinh 2\tilde{\mu})\right]\sqrt{\frac{1-q(\tilde{\mu})^n}{1-q(\tilde{\mu})}}.
\end{equation}
Note that in order to obtain the same thermodynamic relations as \cite{Davison:2016ngz}, we performed a rescaling $\mu \rightarrow \tilde{\mu}=\mu/2$ (see \cref{sec:saddle_appendix}).
All results in the following sections will be given in this scaling of $\mu$. 
\subsubsection{Krylov Complexity in Canonical Ensemble}
We can define the state of the system at time $t$ in a given charge subsector,
$\ket{\psi(\mathcal{Q},t)}$, as being
\begin{equation}\label{eq:Canonical_state_at_time_t}
    \ket{\psi(\mathcal{Q},t)} = \text{e}^{-i t T_H}\ket{0, \mathcal{Q}}.
\end{equation}
The state Krylov complexity defines the expectation value of the number operator.
The number operator in the canonical ensemble is given in \eqref{eq:num_op_Can}.
Thus, we get that,
\begin{align}\label{eq:K_complexity_canonical}
    C^{(\mathcal{Q})}_\mathcal{K}(t) &= \bra{\psi(\mathcal{Q},t)}\hat{n}(t)\ket{\psi(\mathcal{Q},t)} = \sum_n n 
    \braket{\psi(\mathcal{Q},t)\vert n, \mathcal{Q}}\!\!\braket{n, \mathcal{Q}\vert \psi(\mathcal{Q},t)}\nonumber \\ 
    &= \sum_n n \abs{\psi_{n}(\mathcal{Q},t)}^2,
\end{align}
\Cref{eq:K_complexity_canonical} requires us to know an expression for $\psi_n(\mathcal{Q},t)$.
Recall from the discussion in \cref{sec:fixedQTransferMatrix} that the Lanczos coefficients in a fixed 
charge subsector are,
\begin{equation}\label{eq:Lanczos_coeff_Canon}
    b_n(\mathcal{Q}) = \frac{\mathcal{J}(1-4\mathcal{Q}^2)^{\frac{p}{2}}}{\sqrt{\lambda}}\sqrt{\frac{1-q(\mathcal{Q})^n}{1-q(\mathcal{Q})}}\exp\left[-\frac{\lambda}{2}\frac{(4\mathcal{Q}(\mathcal{Q}+1)-1)}{1-4\mathcal{Q}^2}\right]
\end{equation}
which are similar to those obtained in \cite{Berkooz:2020uly} except now the $q$-number along with an overall rescaling are both dependent on the charge of the subsector.
Starting from \eqref{eq:Canonical_state_at_time_t}, we can obtain a Schr\"odinger like equation for the time evolution of the state in the Krylov basis to be,
\begin{align}\label{eq:Schroedinger_like_eq}
    \partial_t\chi_n(\mathcal{Q},t) &= b_n(\mathcal{Q})\chi_{n-1}(\mathcal{Q},t) - b_{n+1}(\mathcal{Q})\chi_{n+1}(\mathcal{Q},t)\nonumber\\
    &= f(\mathcal{Q})\left(\sqrt{\left[n\right]_{q(\mathcal{Q})}}\chi_{n-1}(\mathcal{Q},t) - \sqrt{\left[n+1\right]_{q(\mathcal{Q})}}\chi_{n+1}(\mathcal{Q},t)\right)
\end{align}
where we have redefined $\psi_n(\mathcal{Q},t) = i^n\chi_n(\mathcal{Q},t)$ such that the $\chi_n$ are real 
and have absorbed all the terms independent of $n$ within the Lanczos coefficient into $f(\mathcal{Q})$.
This is similar to the expression obtained in \cite{Caputa:2021sib}.

 Following \cite{Rabinovici:2023yex}, we derive expressions for the Lanczos coefficients
 in the large and small $n$ regimes.
The transition between them occurs when $-n\log q(\mathcal{Q})$ is of order one.
Therefore, the critical value $n_*$ is,
\begin{equation}
    n_* = \frac{1-4\mathcal{Q}^2}{4\lambda} .
\end{equation}
Notably, this is charge dependent.
It is also worth noting that for a fixed $\lambda$ as $\mathcal{Q}$ approaches $\pm 1/2$ the critical value approaches zero.
Hence, the Lanczos coefficient follow the second expression in \eqref{eq:Lanczos_regimes_n} at smaller value of $n$. 
The expressions for $b_n$ in the small and large $n$ regimes are similar to those in the SYK model and are given by,
\begin{align}\label{eq:Lanczos_regimes_n}
   b_n(\mathcal{Q}) \approx
    \begin{cases}
        2\mathcal{J}(1-4\mathcal{Q}^2)^{\frac{p-1}{2}}\sqrt{\frac{n}{1-q(\mathcal{Q})}}\exp\left[-\frac{\lambda}{2}\frac{(4\mathcal{Q}(\mathcal{Q}+1)-1)}{1-4\mathcal{Q}^2}\right], \;\;\; n\lesssim \frac{1-4\mathcal{Q}^2}{4\lambda}\\
        \frac{\mathcal{J}(1-4\mathcal{Q}^2)^{\frac{p}{2}}}{\sqrt{\lambda(1-q(\mathcal{Q}))}}\left(1-\frac{q(\mathcal{Q})^n}{2}\right)\exp\left[-\frac{\lambda}{2}\frac{(4\mathcal{Q}(\mathcal{Q}+1)-1)}{1-4\mathcal{Q}^2}\right], \;\;\; n \gtrsim \frac{1-4\mathcal{Q}^2}{4\lambda}.
    \end{cases}
\end{align}
For $n\rightarrow \infty$, the Lanczos coefficient in a fixed charge subsector approaches a constant,
\begin{equation*}
    b_\infty = \frac{\mathcal{J}(1-4\mathcal{Q}^2)^{\frac{p}{2}}}{\sqrt{\lambda(1-q(\mathcal{Q}))}}\exp\left[-\frac{\lambda}{2}\frac{(4\mathcal{Q}(\mathcal{Q}+1)-1)}{1-4\mathcal{Q}^2}\right] = \text{constant}.
\end{equation*}
For small $\lambda$, \eqref{eq:Lanczos_regimes_n} simplifies to
\begin{align}\label{eq:Lanczos_regimes_lambda}
    b_n(\mathcal{Q}) \approx
    \begin{cases}
        2\mathcal{J}(1-4\mathcal{Q}^2)^{\frac{p-1}{2}}\sqrt{\frac{n}{\lambda}}, \;\;\; 
        n\lesssim \frac{1-4\mathcal{Q}^2}{4\lambda} ,\\
        \frac{\mathcal{J}(1-4\mathcal{Q}^2)^{\frac{p}{2}}}{\sqrt{\lambda(1-q(\mathcal{Q}))}}
        \left(1-\frac{q(\mathcal{Q})^n}{2}\right), \;\;\; n\gtrsim \frac{1-4\mathcal{Q}^2}{4\lambda}.
    \end{cases}
\end{align}
The authors of \cite{Caputa:2021sib} observed that the solution for \eqref{eq:Schroedinger_like_eq} in the case of the Majorana SYK model at early times requires the Lanczos coefficients to be in the small $n$ regime.
In this regime, we recover the cSYK version of the Heisenberg-Weyl algebra in each charge subsector.
Furthermore, as the chord states are coherent states, we can follow the method obtained in \cite{Caputa:2021sib}
and obtain solutions for \eqref{eq:Schroedinger_like_eq} such that now,
\begin{equation}\label{eq:sol_Schroedinger}
    \psi_n(\mathcal{Q},t) = \text{e}^{-\gamma(\mathcal{Q})^2t^2/2}\frac{(-i\gamma(\mathcal{Q})t)^n}{\sqrt{n!}}
\end{equation}
with $\gamma(\mathcal{Q}) = \frac{\mathcal{J}(1-4\mathcal{Q}^2)^\frac{p}{2}}{\sqrt{1-q(\mathcal{Q})}}\exp\left(-\frac{\lambda}{2}\frac{(4\mathcal{Q}(\mathcal{Q}+1)-1)}{1-4\mathcal{Q}^2}\right)$.
Therefore, using \eqref{eq:sol_Schroedinger} in \eqref{eq:K_complexity_canonical} we obtain the Krylov complexity in a given 
charge sector,
\begin{equation}
    C_\mathcal{K}^{(\mathcal{Q})}(t) = (\mathcal{J}t)^2\frac{(1-4\mathcal{Q}^2)^p}{1-q(\mathcal{Q})}\exp\left(-\frac{\lambda(4\mathcal{Q}(\mathcal{Q}+1)-1)}{1-4\mathcal{Q}^2}\right).
\end{equation}
The instance $t_*$ at which a transition in the Krylov complexity due to the transition in the Lanczos coefficients 
occurs can be found via $C_\mathcal{K}^{(\mathcal{Q})}(t_*) = n_*$,
leading to
\begin{equation}
    t_* = \frac{1}{2\mathcal{J}}\exp\left(\frac{\lambda}{2}\frac{(4\mathcal{Q}(\mathcal{Q}+1)-1)}
    {1-4\mathcal{Q}^2}\right)\sqrt{\frac{1-q(\mathcal{Q})}{\lambda(1-4\mathcal{Q}^2)^{p-1}}} . \label{eq:tstar}
\end{equation}
The time $t_*$ corresponds to the scrambling time \cite{Rabinovici:2025otw}.
It diverges as $\mathcal{Q}$ approaches $\pm 1/2$.
For small $\lambda$, the expression \eqref{eq:tstar} simplifies to
\begin{align}
    t_*\approx\frac{1}{\mathcal{J}\left(1-4\mathcal{Q}^2\right)^{p/2}}.
\end{align}
Here, we see that the scrambling time increases with $p$.

In \cite{Barbon:2019wsy}, it was noticed that the solution for \eqref{eq:Schroedinger_like_eq} is given 
by a Bessel function of the first kind, $\psi_n(\mathcal{Q},t) = J_n(2tb_{\infty}(\mathcal{Q}))$.
We obtain for the Krylov complexity in the two regimes 
\begin{equation}
    C_{\mathcal{K}}^{(\mathcal{Q})} = \begin{cases}
        (\mathcal{J}t)^2\frac{(1-4\mathcal{Q}^2)^p}{1-q(\mathcal{Q})}
        \exp\left[-\lambda\frac{(4\mathcal{Q}(\mathcal{Q}+1)-1)}{1-4\mathcal{Q}^2}\right],\;\;\; t\lesssim t_* ,\\
        2\mathcal{J}t\frac{(1-4\mathcal{Q}^2)^{\frac{p}{2}}}{\sqrt{\lambda(1-q(\mathcal{Q}))}}\exp\left[-\frac{\lambda}{2}
        \frac{(4\mathcal{Q}(\mathcal{Q}+1)-1)}{1-4\mathcal{Q}^2}\right],\;\;\; t \gg t_*
    \label{eq:kryltwo}
    \end{cases}
\end{equation}
which simplifies further in the small $\lambda$ limit to be,
\begin{equation}
    C_{\mathcal{K}}^{(\mathcal{Q})} = \begin{cases}
        \frac{(\mathcal{J}t)^2}{4\lambda}(1-4\mathcal{Q}^2)^{p+1}, \;\;\; t\lesssim t_* ,\\
        \frac{\mathcal{J}t(1-4\mathcal{Q}^2)^{\frac{p+1}{2}}}{\lambda},\;\;\; t \gg t_* .
    \end{cases}
\end{equation}

Finally, we solve \cref{eq:Schroedinger_like_eq} numerically to obtain \cref{fig:C_K_multiQ} which shows curves of $C_{\mathcal{K}}^{(\mathcal{Q})}(t)$ for various values of $\mathcal{Q}$.
In \cref{fig:C_K_Q_analytic_vs_numerical}, we compare the approximate analytic solutions \eqref{eq:kryltwo} to the numerically obtained results.
As expected, we get a good fit for early and late times, $t\ll t_\star$ and $t\gg t_\star$ respectively, while around the transition point $t_\star$ the curves slightly differ. 
Furthermore, we also obtain curves for the grand canonical Krylov complexity $C_{\mathcal{K}}(t)$ by again solving the recursion relations numerically, 
but this time using the $\mu$ dependent Lanczos coefficients \eqref{eq:Lanczos_coeff_GC}, they can be found in \cref{fig:C_K_multi_mu}.
We can then integrate \cref{eq:kryltwo} weighted with the exponential of the grand potential $\exp\left[-N\beta \Omega\right]$ as explained in \cref{sec:krylov_in_gc,sec:parameters_appendix} to get an analytic curve for $C_{\mathcal{K}}(t)$.
We compare the analytic to the numeric results in \cref{fig:C_K_mu_analytic_vs_numerical} and find them in good agreement.

	\section{Discussion}
In this paper, we examined the relation between the Krylov complexity of grand canonical and canonical states
of the complex SYK model in its double-scaling limit.
Our major result is the Krylov complexity of the grand canonical ensemble written as a weighted
average of the Krylov complexities in the individual charge sectors in (\ref{eq:GC_Can_Krylov_relation}).
The employed technique was to express the moments of the SYK Hamiltonian as vacuum expectation values in an open oriented chord space. Combinatorial methods presented in \cite{Berkooz:2020uly, Berkooz:SUSY_SYK} were applied. 
We define an auxiliary Hilbert space along with its inner product and the oriented chord creation and annihilation operators.
Combinations of these operators are shown to have the same $q$-deformed commutation relations as the creation and annihilation
operators for unoriented $H$-chords. 
The symmetrized transfer matrix obtained via the oriented chord formalism 
is matched to that for the $H$-chords in both ensembles.
This allowed us to relate the grand canonical to the canonical chord states.
Following this, the relation between the grand canonical  and the canonical Krylov complexities 
was established.
The Krylov complexity was computed in various regimes.\\
The following contains a more detailed account of the results for individual sections after 
section \ref{subsec:GC_oriented_chords}.

\begin{itemize}
    \item  An auxiliary Hilbert 
    space is constructed 
    by acting with  operators $\alpha$ and $\beta$ on the zero chord state. 
    These operators mimic the insertion of a $\psi^I$ or a $\bar{\psi}_I$ in the trace and form a linear combination of oriented
    chord creation and annihilation operators.
   The penalty factors $W_\mu$ and $\tilde{W}_\mu$ associated to closing an $X$ and an $O$ chord, respectively, 
   are given. The creation of a chord is accompanied by multiplying with a $\theta$ and orientation dependent phase factor.

    \item The inner product for the auxiliary Hilbert space provides a way to identify null vectors.
    The physical Hilbert space is obtained by removing the null vectors. 
    It can be split into two sectors. 
    Within each of these sectors there are bosonic states which have equal number of alternating $X$s and $O$s.
    The bosonic states are also accompanied by fermionic states which are  given by alternating $X$s and $O$s as well
    however with the number of $X$s and $O$s differing by one.
    By following \cite{Berkooz:SUSY_SYK} an exact expression for the inner product of these states was derived
    as a solution to a set of linear equations and recursion relations.

    \item Owing to the form of the cSYK Hamiltonian, non-zero contributions to the moments are only
    due to diagrams in which chords are opened or closed in pairs of opposite orientation. 
    These pairs are called $H$-chords \cite{Berkooz:2020uly}.
    The fact that only $H$-chords appear allowed us to simplify the expression for the grand canonical moments and 
    evaluate them via a saddle point analysis as done in \cite{Berkooz:2020uly}.
    The saddle point solution could be used to obtain the penalty factors in a given charge subsector.

    \item The operators $\alpha$ and $\beta$ can be viewed as superpositions of ``pure'' creation or annihilation 
    operators $a,b, a^\dagger \text{ and } b^\dagger$.
    Using these operators we constructed a new set of bosonic operators $A,B$ and $A^\dagger,B^\dagger$ which create and 
    annihilate pairs of chords $XO$ and $OX$ respectively.
    Upon normalizing these operators, a $q$-deformed extended algebra was obtained which matches the extended 
    algebra of the $H$-chord operators.
    This allowed us to identify the symmetrized transfer matrix obtained from the oriented chord formalism with the 
    transfer matrix for the $H$-chords in both ensembles.
    Furthermore, we provide a relation between grand canonical and 
    canonical states with non-zero chord number.

    \item Finally, using the connection between grand canonical and canonical chord states a relation between 
    the corresponding Krylov complexities was obtained. 
For the canonical Krylov complexity we display analytic expressions in the large and small $n$ regimes (separated by 
    the critical value $n_*$). The canonical complexity depends on the charge 
and is maximal for vanishing charge. Numerical calculations confirm our results.
\end{itemize}
For future work, it would be interesting to consider other examples with global symmetries.
One such example is the $\mathcal{N} =2$ supersymmetric SYK.
Our current methods can not be directly applied.
However, our findings suggest that the transfer matrix is not block diagonal i.e. it mixes charge sectors.
Then if true, the weighted average of the canonical Krylov complexities will give a lower bound for the grand canonical complexity.\\
Furthermore, as mentioned in the introduction we plan on applying the methods developed in this paper to investigate the Krylov complexity of matter operators.
Additionally, we would like to relate our results to a bulk interpretation.
Based on our findings, we expect the length of the wormhole in the bulk to grow linearly after scrambling time $t_*$ and to depend on charge/chemical potential.
	\section*{Acknowledgments}

	We thank Vladimir Narovlanski for some useful hints.
	Moreover, we acknowledge support by
	the Bonn Cologne Graduate School of Physics and Astronomy (BCGS) and access to the Marvin cluster of the University of Bonn. 
	Finally, Y.K. would like to thank the Rosa Luxemburg Foundation for the granted financial support. 

	\FloatBarrier
	\newpage
	\begin{appendices}
		\crefalias{section}{appsec}
		\appendix
		\section{Inner Product}
	 		\label{sec:inner_product}
			We will define the inner product on the vector space of chord states.
To this end, we will follow the procedure outlined in \cite{Berkooz:SUSY_SYK}.
For now, we shall work in the subspace with arbitrary but constant auxiliary charge $\theta$.
However, the definition can easily be extended to the entire Fock space/algebra.
Since, in this section we work outside the Fourier integral, we rotate to $i\theta\rightarrow\theta$.
This is important for the inner product to be positive definite.
The inner product is fully determined by demanding
\begin{align}
    \braket{0|0} &\equiv 1, \label{eq:norm}\\
    \beta &\equiv \alpha^\dagger
\end{align}
and requiring that inner products of states with unequal number of $X$s
or $O$s vanish. 
 
Using chord annihilation/creation, we can then build all possible inner products recursively from the zero chord state $\ket{0}$, i.e.
\begin{align}
    &e^{-\theta}\braket{P(n_X, n_O)X|\tilde{P}(n_X, n_O)X} \nonumber\\
    &= \braket{\beta P(n_X, n_O)|\tilde{P}(n_X, n_O)X} \nonumber\\
    &= \braket{P(n_X, n_O)|\alpha \tilde{P}(n_X, n_O)X} \nonumber\\
    &= \bra{P(n_X, n_O)} \left( \sum_{i} W(\tilde{P}(n_X, n_O)X-X_i) \ket{\tilde{P}(n_X, n_O)X-X_i}\right)
    \label{eq:removing_an_x}
\end{align}
and
\begin{align}
    &e^{\theta}\braket{P(n_X, n_O)O|\tilde{P}(n_X, n_O)O} \nonumber\\
    &= \bra{P(n_X, n_O)} \left( \sum_{i} \tilde{W}(\tilde{P}(n_X, n_O)O-O_i) \ket{\tilde{P}(n_X, n_O)O-O_i}\right),
    \label{eq:removing_an_o}
\end{align}

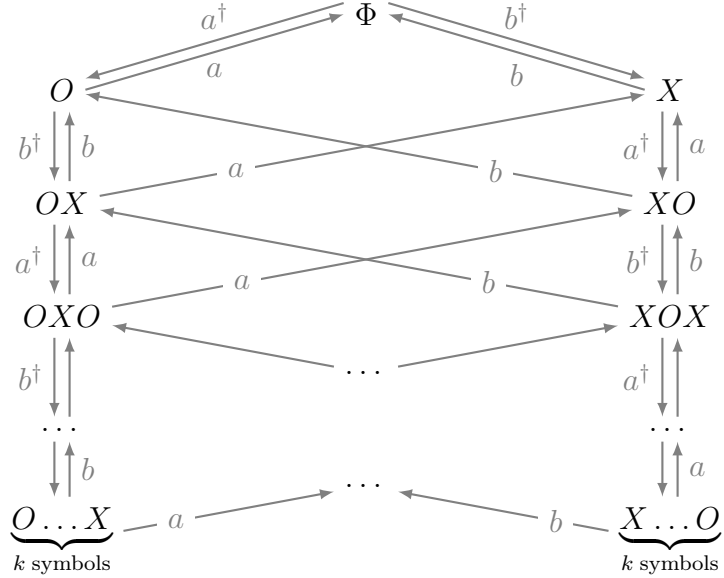
\begin{figure}
\centering
\begin{tikzpicture}
    \node (0) at (0,0) {$\Phi$};

    \node (l1) at (-4,-1) {$O$};
    \node (r1) at (4,-1) {$X$};

    \node (l2) at (-4,-2.5) {$OX$};
    \node (r2) at (4,-2.5) {$XO$};

    \node (l3) at (-4,-4) {$OXO$};
    \node (r3) at (4,-4) {$XOX$};

    \node (l4) at (-4,-5.5) {$\dots$};
    \node (r4) at (4,-5.5) {$\dots$};

    \node (rl4) at (0, -4.75) {$\dots$};

    \node (rl5) at (0, -6.25) {$\dots$};

    \node (l5) at (-4,-7) {$\underbrace{O\dots X}_{k\;\text{symbols}}$};
    \node (r5) at (4,-7) {$\underbrace{X\dots O}_{k\;\text{symbols}}$};

    \draw[gray, thick, -latex] ([yshift=+0.15cm]0.west) -- ([yshift=+0.15cm]l1.east) node [midway, above] {$a^\dagger$};
    \draw[gray, thick, -latex] ([yshift=+0.15cm]0.east) -- ([yshift=+0.15cm]r1.west) node [midway, above] {$b^\dagger$};

    \draw[gray, thick, latex-] (0.west) -- (l1.east) node [midway, below] {$a$};
    \draw[gray, thick, latex-] (0.east) -- (r1.west) node [midway, below] {$b$};

    \draw[gray, thick, -latex] (r2) -- (l1) node [near start, fill=white] {$b$};
    \draw[gray, thick, -latex] (l2) -- (r1) node [near start, fill=white] {$a$};

    \draw[gray, thick, -latex] (r3) -- (l2) node [near start, fill=white] {$b$};
    \draw[gray, thick, -latex] (l3) -- (r2) node [near start, fill=white] {$a$};

    \draw[gray, thick, -latex] (rl4) -- (l3) ;
    \draw[gray, thick, -latex] (rl4) -- (r3) ;

    \draw[gray, thick, -latex] (r5) -- (rl5) node [near start, fill=white] {$b$};
    \draw[gray, thick, -latex] (l5) -- (rl5) node [near start, fill=white] {$a$};

    \draw[gray, thick, -latex] ([xshift=-0.1cm]l1.south) -- ([xshift=-0.1cm]l2.north) node [midway, left] {$b^\dagger$};
    \draw[gray, thick, -latex] ([xshift=0.1cm]l2.north) -- ([xshift=0.1cm]l1.south) node [midway, right] {$b$};

    \draw[gray, thick, -latex] ([xshift=-0.1cm]r1.south) -- ([xshift=-0.1cm]r2.north) node [midway, left] {$a^\dagger$};
    \draw[gray, thick, -latex] ([xshift=0.1cm]r2.north) -- ([xshift=0.1cm]r1.south) node [midway, right] {$a$};

    \draw[gray, thick, -latex] ([xshift=-0.1cm]l2.south) -- ([xshift=-0.1cm]l3.north) node [midway, left] {$a^\dagger$};
    \draw[gray, thick, -latex] ([xshift=0.1cm]l3.north) -- ([xshift=0.1cm]l2.south) node [midway, right] {$a$};

    \draw[gray, thick, -latex] ([xshift=-0.1cm]r2.south) -- ([xshift=-0.1cm]r3.north) node [midway, left] {$b^\dagger$};
    \draw[gray, thick, -latex] ([xshift=0.1cm]r3.north) -- ([xshift=0.1cm]r2.south) node [midway, right] {$b$};

    \draw[gray, thick, -latex] ([xshift=-0.1cm]l3.south) -- ([xshift=-0.1cm]l4.north) node [midway, left] {$b^\dagger$};
    \draw[gray, thick, -latex] ([xshift=0.1cm]l4.north) -- ([xshift=0.1cm]l3.south);

    \draw[gray, thick, -latex] ([xshift=-0.1cm]r3.south) -- ([xshift=-0.1cm]r4.north) node [midway, left] {$a^\dagger$};
    \draw[gray, thick, -latex] ([xshift=0.1cm]r4.north) -- ([xshift=0.1cm]r3.south);

    \draw[gray, thick, -latex] ([xshift=-0.1cm]l4.south) -- ([xshift=-0.1cm]l5.north);
    \draw[gray, thick, -latex] ([xshift=0.1cm]l5.north) -- ([xshift=0.1cm]l4.south) node [midway, right] {$b$};

    \draw[gray, thick, -latex] ([xshift=-0.1cm]r4.south) -- ([xshift=-0.1cm]r5.north);
    \draw[gray, thick, -latex] ([xshift=0.1cm]r5.north) -- ([xshift=0.1cm]r4.south) node [midway, right] {$a$};

\end{tikzpicture}
\caption{States in the physical Hilbert space that contribute to the $k$th moment. The operators $a$ and $b$ are defined in \cref{sec:bosonic_sector}.}
\label{fig:physical_hilbert_space}
\end{figure}

From (\ref{eq:expression_for_W}) and (\ref{eq:expression_for_tilde_W}), it can be seen that penalty factors for closing adjacent $X$ 
respectively $O$ chords only differ by a sign if $p$ is odd
\begin{align}
   W\left(\dots \wick{\c1XX \dots \c1O}\right) &= -W\left(\dots \wick{X\c1X \dots \c1O}\right) , \\
   \tilde{W}\left(\dots \wick{\c1OO \dots \c1X}\right) &= -\tilde{W}\left(\dots \wick{O\c1O \dots \c1X}\right) .
\end{align}
Consequently, in that case any vector with a non-alternating string of $X$s and $O$s must be identified with the null vector $\vec{0}$,
in order to have a consistently defined Hilbert space.
In \cite{Berkooz:SUSY_SYK}, this problem is solved by fixing a charge sector and then modding out the vectors with non-alternating strings.
In the case of the complex SYK, each Hamiltonian insertion represents a pair of $\Psi$ and $\Psi^\dagger$ and thus $X$/$O$-strings will be naturally alternating.
\Cref{fig:physical_hilbert_space} shows the remaining states that contribute to the $k$th moment.

Let $n$ denote a string of $n$ $XO$-pairs and $\overline{n}$ a string of $OX$-pairs. 
For every $n\in \mathbb{N}$, there are four possible states in the physical Hilbert space, 
the two bosonic states, $\ket{n}$ and $\ket{\overline{n}}$, and the two fermionic states, $\ket{nX}$ and $\ket{\overline{n}O}$. 
This is analogous to \cite{Berkooz:SUSY_SYK}, even though we changed the notation slightly. 
Still following \cite{Berkooz:SUSY_SYK}, we can find a general expression for the inner product on the physical Hilbert space via recursion.
To that end, we first define
\begin{equation}
    \begin{aligned}[c]
    \braket{n,n}&\equiv A_n &\braket{\overline{n},\overline{n}}&\equiv B_n\\
    \braket{nX,nX}&\equiv C_n &\braket{\overline{n}O,\overline{n}O}&\equiv D_n\\
    \mathrlap{\braket{n,\overline{n}}\equiv E_n}
    \end{aligned}
\end{equation}
and then generate a system of linear equations via \eqref{eq:removing_an_x} and \eqref{eq:removing_an_o}.
Solving it gives the following recursion relations, for $n\geq 1$
\footnote{
    For $n\ge1$, acting with $a$ on $\ket{n+1}$ results in the fermionic state $\ket{\overline{n}X}$, from where acting with $b$ either takes us to $\ket{n}$ or to $\ket{\overline{n}}$.
    We thus get two contributions to $ba\ket{n+1}$.
    Contrast this with the $n=0$ case. 
    Here, we arrive at $X$ after initially acting with $a$ $\ket{1}$, but from there we only have one way to go down to the zero chord state.
    For this reason, \cref{eq:recursion_An} only holds for $n\geq 1$.
    Similar arguments hold for \cref{eq:recursion_Bn}.
} 
we have
\begin{align}
    \exp\left(-2\lambda e^{\mu}\cosh\mu \right)A_{n+1} &= \left(1-\exp\left[-4\lambda n\cosh^2\mu \right]\right)A_n,\label{eq:recursion_An}\\
    \exp\left(-2\lambda e^{-\mu}\cosh\mu \right)B_{n+1} &= \left(1-\exp\left[-4\lambda n\cosh^2\mu \right]\right)B_n\label{eq:recursion_Bn},
\end{align}
with all other symbols related to $A_n$ and $B_n$ via 
\begin{align}
    C_n &= A_n e^{\theta}\left[1-\exp\left(-4n\lambda \cosh^2\mu \right)\right],\\
    D_n &= B_n e^{-\theta}\left[1-\exp\left(-4n\lambda \cosh^2\mu \right)\right],\\
    E_n &= -A_n \exp \left( -2n\lambda e^\mu \cosh\mu \right),\\
    E_n^{*} &= -B_n \exp \left( -2n\lambda e^{-\mu} \cosh\mu \right).
\end{align}
For all $n\ge 1$, \cref{eq:recursion_An,eq:recursion_Bn} are solved by
\begin{align}
    &\begin{aligned}
     A_n &= (-1)^{n-1} \exp\left(\lambda (n-1)\left[1+e^{2\mu}-2n\cosh^2\mu \right]\right)\exp\left(2\lambda e^{\mu}\cosh\mu \right)\\
     &\hphantom{=}\times\left(e^{4\lambda\cosh^2\mu};\,e^{4\lambda\cosh^2\mu}\right)_{n-1},
    \end{aligned}\label{eq:inner_product_bosonic}\\
  &\begin{aligned}
     B_n &= (-1)^{n-1} \exp\left(\lambda (n-1)\left[1+e^{-2\mu}-2n\cosh^2\mu \right]\right)\exp\left(2\lambda e^{-\mu}\cosh\mu \right)\\
     &\hphantom{=}\times\left(e^{4\lambda\cosh^2\mu};\,e^{4\lambda\cosh^2\mu}\right)_{n-1},
    \end{aligned}
\end{align}
where $(a;\, q)_n$ is the $q$-Pochhammer symbol, and we used \cref{eq:norm} to fix the respective constant in the general solutions.
Notice that the $q$-Pochhammer symbols in the equations above alternate between negative and positive values, 
thus together with the alternating factor of $(-1)^{n-1}$ in front of each equation the inner product is ultimately positive-definite.
Finally, using $A_n(B_n)$, we can introduce the normalized states 
\begin{align}\label{eq:norm_states}
    &\ket{\tilde{n}} = \frac{\ket{n}}{\sqrt{A_n}}, & \ket{\bar{\tilde{n}}} = \frac{\ket{\bar{n}}}{\sqrt{B_n}},
\end{align}
for all $n\ge 1$ (the zero chord state is taken to be normalized to one).
		\section{Saddle Point Approximation}
			\label{sec:saddle_appendix}
At several points throughout the paper, we solve the integral
\begin{align}
    m_k(Q)\propto\int_{-\infty}^\infty d\mu e^{-i\mu Q} m_k(i\mu)\nonumber
\end{align}
via saddle point approximation in the large $N$-limit. 
This has been done equivalently in \cite{Berkooz:2020uly} and similarly in \cite{Berkooz:SUSY_SYK}.
For clarity of notation, we will repeat the calculation here.

We begin by nominally introducing rescaled versions of $\mu$ and $Q$,
$\tilde{\mu}\equiv\mu/2$ and $\tilde{Q}\equiv 2Q$. 
This is only to ensure that our $\mu$ and $Q$ scale the same as in \cite{Davison:2016ngz}, and we can compare our results 
(our $\mu$ actually corresponds to $\beta \mu$ in \cite{Davison:2016ngz}).
We also introduce a normalized version of the charge $Q\equiv N\mathcal{Q}$ 
and demand $\mathcal{Q}=\text{const.}$ as $N\rightarrow \infty$, 
again using the notation of \cite{Davison:2016ngz}.
We can then write
\begin{align}
    m_k(Q) &= \frac{1}{2\pi} \int_{-\infty}^\infty d\tilde{\mu} e^{-i\tilde{\mu} \tilde{Q}} m_k(i\tilde{\mu}) \nonumber\\
        &=\frac{1}{4\pi} \int_{-\infty}^\infty d\mu e^{-iN\mu \mathcal{Q}} m_k(i\mu/2).
\end{align} 
Using the definition of $m_k(\mu)$ \eqref{eq:GC_moments}, we group all terms in the exponent into three functions.
We collect the $N$-dependent terms in
\begin{align}
    N F(\mu) = N\left[\left(1-k\sqrt{\frac{\lambda}{N}}\right)\ln \left(\cos\left[\frac{\mu}{2}\right]\right)+i\mu\mathcal{Q}\right],
\end{align}
the $m$ dependent terms in
\begin{align}
    g(\mu)=\frac{i\mu m}{2} - i\lambda\frac{m}{2}(k-1)\sin{\mu}+\lambda\frac{m^2}{2}\cos^2\left(\frac{\mu}{2}\right)
\end{align}
and the remaining terms in 
\begin{align}
    f(\mu) = \lambda\frac{k}{2}(1-i\sin(\mu))-\lambda\frac{k^2}{2}\sin^2\left(\frac{\mu}{2}\right).
\end{align}
As we send $N\rightarrow\infty$, $NF(\mu)$ dominates over all other contributions, 
and we are therefore justified in using Laplace's method to approximate the integral.
Since $F(\mu)$ is extremal at 
\begin{align}
    \mu_0\equiv-2\arctan\left(\frac{i 2 \mathcal{Q}}{1-k\sqrt{\frac{\lambda}{N}}}\right),\label{eq:mu_0}
\end{align}
the integral approaches
\begin{align}
    m_k(\mathcal{Q}) &=\sum_{m}\frac{\mathcal{J}^k}{\lambda^\frac{k}{2}}\frac{1}{4\pi} \int_{-\infty}^\infty d\mu \exp\left[NF(\mu)+g(\mu)+f(\mu)\right] m_{k;m}(\mu/2) \\
    &\approx \frac{\mathcal{J}^k}{\lambda^\frac{k}{2}} \sqrt{\frac{1}{8\pi N F^{\prime \prime}(\mu_0)}} \exp\left[NF(\mu_0)+g(\mu_0)+f(\mu_0)\right] m_{k;m}(\mu_0/2).
\end{align}
Here, to zeroth order in $N^{-1/2}$ we find
\begin{align}
   N F(\mu_0) &=-2N\mathcal{Q}\artanh(2\mathcal{Q})+\frac{1}{2}\left[k\sqrt{\lambda N}-N\right]\log\left(1-4\mathcal{Q}^2\right)-\frac{2k^2 \mathcal{Q}^2\lambda}{1-4\mathcal{Q}^2},\\
   N F^{\prime\prime}(\mu_0) &=-\frac{N}{4} \left(1-4\mathcal{Q}^2\right)+\frac{1}{4} k \sqrt{\lambda N} \left(4 \mathcal{Q}^2+1\right),\\
   f(\mu_0) &=-\frac{k \lambda  (4 \mathcal{Q} (\mathcal{Q}+1)-1)}{2\left(1-4 \mathcal{Q}^2\right)} +\frac{2 k^2 \mathcal{Q}^2 \lambda}{1-4 \mathcal{Q}^2},\\
   g(\mu_0) &=\frac{\lambda  m (m-4 (k-1) \mathcal{Q})}{2\left(1-4 \mathcal{Q}^2\right)}+2 m \tanh ^{-1}(2 \mathcal{Q}).
\end{align}
Moreover, we manually insert a factor of $2^N$ to properly normalize the trace ($\tr(\mathbb{I})\equiv 1$).
In total, we find
\begin{align}
    m_k(\mathcal{Q})
    &\approx \frac{\mathcal{J}^k}{\lambda^\frac{k}{2}} \exp\left[
        N\left(\!-\frac{1}{2} \log \left(\frac{1-4\mathcal{Q}^2}{4}\right)-2\mathcal{Q} \artanh(2\mathcal{Q})\right)
        +\frac{1}{2} k \sqrt{\lambda N } \log \left(1-4\mathcal{Q}^2\right)
   \right]\nonumber\\
    &\hphantom{=}\times\sqrt{\frac{1}{8\pi N\left(1-4\mathcal{Q}^2\right)}}\exp\left[
        -\frac{k \lambda  (4\mathcal{Q} (\mathcal{Q}+1)-1)}{2 \left(1-4\mathcal{Q}^2\right)}
        \right]\label{eq:m_of_Q} \\
    &\hphantom{=}\times\sum_{m=-k,\cdots,k}\exp\left[
        m \left(\artanh(2\mathcal{Q})+\frac{\lambda  (m-2 (k-1) 2\mathcal{Q})}{2 \left(1-4\mathcal{Q}^2\right)}\right)
        \right]m_{k;m}(\mu_0/2). \nonumber 
\end{align}
		\section{Parameter Dependence of Operators and States}
			\label{sec:parameters_appendix}
In this appendix, we will discuss the dependence of operators and states on the parameters $m$, $\theta$, $\mu$ and $Q$.
We will show, that we can work in two equivalent pictures where either states or operators depend on the parameters above.
Let us choose the auxiliary charge $\theta$ as our starting point.

In \cref{sec:moments}, we introduced the transfer matrix $T(\theta)$ on the space of the auxiliary charge $\theta$. 
Although, at this point the form of the transfer matrix is not clear, 
we already know that it somehow has to depend on the subsequently introduced operators $\alpha$ and $\beta$.
Accordingly, these should be $\theta$ dependent too.
The states that these operators act on were consequently defined to be $\theta$ independent.
\begin{align}
    (\alpha + \beta)(\theta) \ket{0} = e^{-i\theta}\ket{X} + e^{i\theta}\ket{O}
\end{align}
Nevertheless, we could equivalently have defined the $\theta$-dependence to reside in the states, like so
\footnote{Here, we switch from states in the Hilbert spaces with fixed auxiliary charge $\theta$, $\ket{n}\in\mathcal{H}_\theta$, 
to states in the Fock space of all charges, $\ket{n,\theta}\in\bigoplus_\theta \mathcal{H}_\theta$.}
\begin{align}
    (\alpha + \beta) \ket{0, \theta} = e^{-i\hat{\theta}}\ket{X, \theta} + e^{i\hat{\theta}}\ket{O, \theta},
\end{align}
where for consistency we require two states with different value of $\theta$ to be orthogonal under our existing inner product definition (see \cref{sec:inner_product}),
i.e. $\braket{\dots,\theta|\dots,\theta^\prime}\propto\delta_{\theta,\theta^\prime}$.
We shall now work out what effect this redefinition has on the subsequent argumentation.

The moment for fixed $k$ and $m$ can be written in terms of the $\theta$-dependent states as
\begin{align}
     m_{k;m}=\left(\int_0^{2\pi}\frac{d\theta}{\sqrt{2\pi}}  \bra{0,
     \theta} \right)  
     T^k_\mu\left( i \hat{\theta}\right)
     \left(\int_0^{2\pi}\frac{d\theta^\prime}{\sqrt{2\pi}} e^{-im\theta^\prime} \ket{0, \theta^\prime}\right). \label{eq:m_km_appendix}
\end{align}  
We can then introduce states with constant $m$
\begin{align}
    \ket{0,m}&=\int_0^{2\pi}\frac{d\theta^\prime}{\sqrt{2\pi}}
    e^{-i{m}\theta^\prime} \ket{0, \theta^\prime} .
\end{align}
Acting with $\exp\left(\mp
in\hat{\theta}\right)$ shifts the value of $m$
by $\pm n$,
\begin{align}
    e^{\mp in\hat{\theta}}\ket{0,m} = \ket{0,(m\pm n)}.
\end{align}
Using our newly defined states, we can now rewrite \cref{eq:m_km_appendix} as
\begin{align}
\label{eq:b6}
    m_{k;m}=\bra{0,0}T^k _\mu\left( i \hat{\theta}\right)\ket{0,m}.
\end{align}

The moment as a function of the $U(1)$ charge, $Q$, is obtained as
a Fourier transform
\begin{equation}
    m_k(Q) =\frac{1}{4\pi} \int_{-\infty}^\infty d\mu e^{-i\mu Q}
    m_k(i\mu/2) ,
\end{equation}
where $m_k\left( \mu\right)$ is defined in
(\ref{eq:GC_moments_as_subspace_sum}). 
We can solve the $\mu$ integral by doing a saddle point approximation.
The corresponding calculation can be found in \cref{sec:saddle_appendix}.
We express $m_{k;m}\left(\mu_0\right)$ in (\ref{eq:m_of_Q}) in terms of a transfer matrix
$T_{\mu_0}\left( i \hat{\theta}\right)$. 
In the following, we will drop the subscript on $T$ whenever its value is given by the saddle point value, $T\equiv T_{\mu_0}$. 
Also notice that in the complex DSSYK model chords are opened or closed in pairs of opposite orientation. 
Hence, the transfer matrix preserves $m$ and thus the dependence on $\hat{\theta}$ cancels, see section \ref{sec:Compare_To_GC}. 
\Cref{eq:b6} then implies that for even $k$ only $m_{k;0}$ is non-vanishing.
Therefore, $m_k(\mathcal{Q})$ reduces to
\begin{equation}
\begin{aligned}
   m_k(\mathcal{Q})&=\frac{\mathcal{J}^k}{\lambda^\frac{k}{2}} \sqrt{\frac{1}{8\pi N\left(1-4\mathcal{Q}^2\right)}} \\
   &\hphantom{=}\times\exp\bigg[
        -N\beta F(\mathcal{Q,\beta})
       +k \left(
        -\frac{ \lambda  (4\mathcal{Q} (\mathcal{Q}+1)-1)}{2 \left(1-4\mathcal{Q}^2\right)}
        +\sqrt{\lambda N }\frac{1}{2}\log\left(1-4\mathcal{Q}^2\right)
    \right)
    \bigg]\\
    &\hphantom{=}\times\bra{0}T^k\ket{0}.
\end{aligned}
\end{equation}
Here,
\begin{align}
    F(\mathcal{Q},\beta) =-\frac{1}{\beta} \left[
        \frac{1}{2}\ln\left(\frac{4}{1-4\mathcal{Q}^2}\right)
        +\mathcal{Q}\ln\left(\frac{1-2\mathcal{Q}}{1+2\mathcal{Q}}\right)
        \right] + \mathcal{O}(p^{-1}) \label{eq:Free_energy}
\end{align}
is the free energy of the complex SYK model in the large $p$ limit \cite{Davison:2016ngz}.
We can now proceed as in the case of the auxiliary charge.
First, we promote the charge to an operator $\hat{Q}$ and assign an
eigenvalue $Q$ to the state. In the expressions for the penalty factors
we replace $Q$ by $\hat{Q}$.
We then perform an inverse Fourier transform, 
\begin{align}
    m_k(\mu) &= \frac{\mathcal{J}^k}{\lambda^\frac{k}{2}}\sqrt{\frac{N}{2\pi}} \int_{-1/2}^{1/2}d\mathcal{Q}  \bra{0,\mathcal{Q}} \exp\left[
        -N\beta\Omega(i\mathcal{Q}, \beta)+\dots
        \right] T^k(i\hat{\mathcal{Q}})\ket{0,\mathcal{Q}}\label{eq:m_k_mu_fourier}\\
    &\approx \sqrt{\frac{N}{2\pi}}\exp\left[-N\beta \Omega(i\mathcal{Q}_0)\right]\nonumber\\
    &\;\;\;\;\times \int_{-1/2}^{1/2}d\mathcal{Q}\bra{0,\mathcal{Q}_0} \exp\left[
       -\frac{N}{2}\beta \Omega^{\prime\prime}(i\mathcal{Q}_0) \left(\mathcal{Q}-\mathcal{Q}_0\right)^2
    \right]T^k(i\hat{\mathcal{Q}_0})\ket{0,\mathcal{Q}_0},\label{eq:m_k_mu_fourier_saddle}
\end{align}
where the dots represent terms sub-leading in $N$, and we identified the grand potential 
\begin{align}
    \Omega(\mu, \beta) = F(\mathcal{Q},\beta)-\frac{\mu}{\beta}\mathcal{Q} \label{eq:grand_potential}
\end{align}
(notice that compared to \cite{Davison:2016ngz} we have $\mu/\beta$ instead of just $\mu$,
this is due to \cref{eq:K} which amounts to a rescaling of the usual $\mu$). 
The primes indicate the derivative w.r.t $\mathcal{Q}$.
In \eqref{eq:m_k_mu_fourier_saddle}, we have absorbed all terms linear in $k$ into the definition of the transfer matrix.
Notice the term odd in $\mathcal{Q}\rightarrow-\mathcal{Q}$.
After $\mathcal{Q}\rightarrow i\mathcal{Q}$, this becomes a phase and should be absorbed in re-definition of the states.
In the second line, we minimized the grand potential and approximated the integral via its saddle at
\begin{align}
    \mathcal{Q}_0=\frac{1}{2}\tan\left[\frac{i\mu}{2}\right].\label{eq:Q_0}
\end{align}
The penalty factors for closing a chord do not depend on $N$ and therefore have no influence on the saddle.
The integral in \eqref{eq:m_k_mu_fourier_saddle} evaluates to,
\begin{equation}
    \int_{-1/2}^{1/2}\dd{\mathcal{Q}}\exp\left[-\frac{N}{2}\beta \Omega^{\prime\prime}(i\mathcal{Q}_0) \left(\mathcal{Q}-\mathcal{Q}_0\right)^2\right] = \frac{4}{\sqrt{\pi}}\Im{\text{erf}\left(\frac{i-\tanh\mu}{4N\cosh^2(\mu/2)}\right)}
\end{equation}
We are therefore justified in introducing the grand canonical states
\begin{align}
    \ket{n,\mu}\equiv \frac{1}{\sqrt{\mathcal{N}}}\int_{-1/2}^{1/2}d\mathcal{Q}e^{-\frac{N}{2}\beta\Omega(\mu,\mathcal{Q})}\ket{n,\mathcal{Q}}.
\end{align}
Notice, that the integral in \cref{eq:m_k_mu_fourier} evaluates to a real value after performing the saddle point analysis and therefore at large $N$ the bra state can be defined in the same fashion.
The reader should also keep in mind that these states do not form an orthonormal basis of the Fock space. 
We choose the normalization factor $\mathcal{N}$ s.t. $\braket{0,\mathcal{Q}| 0,\mathcal{Q}}=1 = \braket{0,\mu| 0, \mu}$.
Thus,
\begin{align}
    \mathcal{N} = \sqrt{\pi}e^{N\log\left[2\cosh\left(\frac{\mu}{2}\right)\right]}\Im{\text{erf}\left(\frac{i-\tanh\mu}{4N\cosh^2(\mu/2)}\right)}. \label{eq:normalization}
\end{align}
Finally, again using that the transfer matrix is block diagonal in $Q$, we write
\begin{align}
    m_k(\mu)\propto \sqrt{\frac{N}{2\pi}} \bra{0,\mu} T^k \ket{0,\mu}.
\end{align}
When acting with the transfer matrix on the grand canonical states, we collect penalties according to \cref{eq:expression_for_W,eq:expression_for_tilde_W}.

As a last remark, notice that the function we extremised to obtain the saddle of the fixed $Q$ ensemble,
or in other words the canonical ensemble, was \cref{eq:Free_energy}, the free energy.
To go back to the grand canonical ensemble, we perform the Legendre transform (\ref{eq:grand_potential}).
Now, the saddle points are located at extrema of the grand potential. 
The positions of the respective saddle in $\mu$ \eqref{eq:mu_0} and $\mathcal{Q}$ \eqref{eq:Q_0}  
correspond to the first order expressions for mean charge and chemical
potential in the large $p$-limit obtained in \cite{Davison:2016ngz}.

		\section{Notes on Charge Mixing}
			\label{app:charge_mixing}
In this appendix, we will generalize from the discussion in the present paper and consider a simplified $U(1)$-symmetric model featuring an oriented chord description.
Specifically, we will comment on potential off-diagonal blocks in the effective grand canonical transfer matrix in the $(n,Q)$-basis, where $Q$ is the $U(1)$-charge.
In the chord description of the canonical ensemble, we stipulate that moments of the Hamiltonian of said model can be written as
\begin{align}
    m_k(Q) = \bra{0}T_Q^k \ket{0},
\end{align}
with
\begin{align}
    T_Q=a+a^\dagger +b + b^\dagger,
\end{align}
where the action of $a^\dagger$($b^\dagger$) is to insert an $O$($X$) with a phase $\exp(i Q)$($\exp(-i Q)$)
and the action of $a$($b$) is to delete an $O$($X$) from the string with some penalty,
similar to the set-up studied in this paper and to the SUSY SYK version \cite{Berkooz:SUSY_SYK}.
The inner product on the space of chord-states is defined analogously to \cref{sec:inner_product}.
For simplicity, we will again assume the penalties to be such that states with non-alternating strings of $X$/$O$s drop out.
Since as usual moments of odd order vanish, we can restrict our discussion to bosonic states $\{\ket{n}, \ket{\bar{n}}\}$,
equivalent to those defined in \cref{sec:inner_product}.

Next, we introduce a chemical potential $\mu$ and define grand-canonical moments,
\begin{align}
    m_k(\mu) = \sum_Q\, e^{i\mu Q}\bra{0}T_Q^k \ket{0}. \label{app_eq:mu_moment}
\end{align}
Note that the block structure is required by $U(1)$ symmetry.
From here on, we will denote by $T_{\hat{Q}}$ the block-diagonal transfer matrix on the Fock-space $\mathcal{F}\equiv \bigoplus_Q H_Q$, spanned by $\{\ket{n,Q}\}$, 
with inner product $\braket{n, Q | n^\prime, Q^\prime}\propto \delta_{n, n^\prime} \delta_{Q, Q^\prime}$, 
each block being the canonical transfer matrix for fixed charge $Q$, i.e.
\begin{align}
    m_{2k}(\mu) = \sum_{Q, Q^\prime} \, \bra{0, Q}T_{\hat{Q}}^{2k} e^{-i\mu \hat{Q}} \ket{0, Q^\prime}.\label{app_eq:gc_moments}
\end{align}
Notice that for an equivalent effective description (without changing the moments), 
we can replace $T_{\hat{Q}}^{2k} e^{-i\mu Q}$ by $T_{\hat{\mu}}^k$ which includes off-diagonal blocks
\begin{align}
    T_\mu^k = \left(
    \begin{matrix}
        \ddots &  &  &\\
        \dots & T_Q^{2k} e^{-i\mu Q} & A & \dots \\
        \dots & B & T_{Q+1}^{2k}e^{-i\mu Q}  & \dots \\
        &&& \ddots
    \end{matrix}
    \right),
\end{align}
with the condition that $\bra{0, Q^\prime}A \ket{0, Q} = \bra{0, Q^\prime}B \ket{0, Q} = 0$.
We shall refer to $T_{\mu}$ as the (effective) grand canonical transfer matrix, 
by which we mean that up to some cut-off $\tilde{k}$ there is an effective description
\begin{align}
    \forall k< \tilde{k}: m_{2k}(\mu) = \sum_{n<k/2} c_n \braket{n,\mu| n,\mu} = \sum_{Q, Q^\prime} \bra{0,Q}T_\mu^k \ket{0,Q}.
\end{align}
It may not always be possible to find such a matrix for infinite $\tilde{k}$.
Although we made it seem like $T_{\mu}^k$ is just a more general version of $T_{\hat{Q}}^{2k}$, we are not free in our choice of the off-diagonal blocks ($A$, $B$, $\dots$).
More precisely, they are determined by our choice of $\mu$.

Due to the factor of $e^{-i\mu \hat{Q}}$ in \cref{app_eq:gc_moments}, 
only those terms survive that come with the correct phase, $e^{i\mu \hat{Q}}$, to cancel it.
Acting repeatedly with $T_{\hat{Q}}$ on the state $\sum_Q \ket{0, Q}$,
and then removing the states with incorrect phase, we project onto the subspace 
\begin{align}
H_{\mu}=\vecspan\left\{\sum_{n, Q^\prime} \ketbra{n, Q^\prime } e^{-i\mu Q^\prime} T_Q^{2k} \sum_Q \ket{0,Q}\Big\vert \; \forall \;k\in \mathbb{N}\cup \{0\}\right\}.
\end{align}
We define the operator $T_{\mu}^k$ as the projection of $T_{\hat{Q}}^{2k}$ onto this subspace.
We may want to introduce a basis ${\ket{n, \mu}}$ on this subspace,
such that $T_\mu \equiv h + h^\dagger$, with
\begin{align}
    h \ket{n, \mu} &\propto \ket{n-1, \mu}, \\
    h^\dagger \ket{n, \mu} &\propto \ket{n+1, \mu}
\end{align}
and the seed state being
\begin{align}
    \ket{0,\mu} \equiv \sum_Q = \ket{0, Q}.
\end{align}
This may not always be possible to solve,
but we can at least get an effective version by imposing a cut-off on $k$.
Notice that we have not performed a mere basis change from the $(n, Q)$-basis here,
since we are simultaneously performing a projection onto $H_\mu$ and therefore $T_\mu$ and $T_{\hat{Q}}$ may not be identical.
Moreover, $\ket{n, \mu}$ with $n>0$ in general will include non-trivial combinations of different chord numbers in the $\ket{n,Q}$-basis,
that is
\begin{align}
\ket{n,\mu}= \sum_{n^\prime\le n, Q} f(n^\prime,Q) \ket{n, Q}.
\end{align}
It is clear that iff $\ket{n,\mu}= \sum_{ Q} e^{i\mu Q} \ket{n, Q}$, the off-diagonal terms in the grand canonical transfer matrix vanish and $T_\mu \propto T_Q$.
Finally, we can again go to a Fock space version $\tilde{\mathcal{F}}\equiv \bigoplus_\mu H_\mu$
\footnote{$\tilde{\mathcal{F}} \subseteq \mathcal{F}$, since due to $U(1)$, symmetry acting with $T_{\hat{Q}}$ does not necessarily explore the entire Fock space.}, 
by promoting $\mu$ to an operator.
Then, the matrix $T_{\hat{\mu}}$ is block diagonal on this space with the blocks being the $T_\mu$s, while clearly $T_{\hat{Q}}$ gets off-diagonal blocks due to the phases from opening chords.
Conversely, when opening/closing chords, $T_\mu$ will introduce phases $\exp(\pm i\hat{\mu})$, which when acting on the charge-eigenstates will cause a shift in $Q$.
In the charge basis, these phases will introduce off-diagonal terms in the matrix form of $T_{\hat{\mu}}$.
On the diagonal, they combine to the proportionality factor of $\exp(-i\mu Q)$, between $T_{\hat{\mu}}$ and $T_{\hat{Q}}$.
	\end{appendices}
	\FloatBarrier
	\newpage
	
\bibliographystyle{utphys.bst}
\bibliography{literature}

@article{Berkooz:SUSY_SYK,
    author = "Berkooz, Micha and Brukner, Nadav and Narovlansky, Vladimir and Raz, Amir",
    title = "{The double scaled limit of Super--Symmetric SYK models}",
    eprint = "2003.04405",
    archivePrefix = "arXiv",
    primaryClass = "hep-th",
    doi = "10.1007/JHEP12(2020)110",
    journal = "JHEP",
    volume = "12",
    pages = "110",
    year = "2020"
}

@article{Berkooz:2020uly,
    author = "Berkooz, Micha and Narovlansky, Vladimir and Raj, Himanshu",
    title = "{Complex Sachdev-Ye-Kitaev model in the double scaling limit}",
    eprint = "2006.13983",
    archivePrefix = "arXiv",
    primaryClass = "hep-th",
    doi = "10.1007/JHEP02(2021)113",
    journal = "JHEP",
    volume = "02",
    pages = "113",
    year = "2021"
}

@article{Rabinovici:2023yex,
    author = "Rabinovici, E. and S{\'a}nchez-Garrido, A. and Shir, R. and Sonner, J.",
    title = "{A bulk manifestation of Krylov complexity}",
    eprint = "2305.04355",
    archivePrefix = "arXiv",
    primaryClass = "hep-th",
    doi = "10.1007/JHEP08(2023)213",
    journal = "JHEP",
    volume = "08",
    pages = "213",
    year = "2023"
}

@article{Ambrosini:2024sre,
    author = "Ambrosini, Marco and Rabinovici, Eliezer and S{\'a}nchez-Garrido, Adri{\'a}n and Shir, Ruth and Sonner, Julian",
    title = "{Operator K-complexity in DSSYK: Krylov complexity equals bulk length}",
    eprint = "2412.15318",
    archivePrefix = "arXiv",
    primaryClass = "hep-th",
    reportNumber = "CERN-TH-2025-040",
    month = "12",
    year = "2024"
}

@article{Caputa:2025mii,
    author = "Caputa, Pawel and Di Giulio, Giuseppe and Loc, Tran Quang",
    title = "{Growth of block diagonal operators and symmetry-resolved Krylov complexity}",
    eprint = "2507.02033",
    archivePrefix = "arXiv",
    primaryClass = "hep-th",
    reportNumber = "YITP-25-101",
    month = "7",
    year = "2025"
}

@article{Gu:2019jub,
    author = "Gu, Yingfei and Kitaev, Alexei and Sachdev, Subir and Tarnopolsky, Grigory",
    title = "{Notes on the complex Sachdev-Ye-Kitaev model}",
    eprint = "1910.14099",
    archivePrefix = "arXiv",
    primaryClass = "hep-th",
    doi = "10.1007/JHEP02(2020)157",
    journal = "JHEP",
    volume = "02",
    pages = "157",
    year = "2020"
}

@article{Davison:2016ngz,
    author = "Davison, Richard A. and Fu, Wenbo and Georges, Antoine and Gu, Yingfei and Jensen, Kristan and Sachdev, Subir",
    title = "{Thermoelectric transport in disordered metals without quasiparticles: The Sachdev-Ye-Kitaev models and holography}",
    eprint = "1612.00849",
    archivePrefix = "arXiv",
    primaryClass = "cond-mat.str-el",
    doi = "10.1103/PhysRevB.95.155131",
    journal = "Phys. Rev. B",
    volume = "95",
    number = "15",
    pages = "155131",
    year = "2017"
}

@article{Bhattacharya:2017vaz,
    author = "Bhattacharya, Ritabrata and Chakrabarti, Subhroneel and Jatkar, Dileep P. and Kundu, Arnab",
    title = "{SYK Model, Chaos and Conserved Charge}",
    eprint = "1709.07613",
    archivePrefix = "arXiv",
    primaryClass = "hep-th",
    doi = "10.1007/JHEP11(2017)180",
    journal = "JHEP",
    volume = "11",
    pages = "180",
    year = "2017"
}

@article{Rabinovici:2025otw,
    author = "Rabinovici, Eliezer and S{\'a}nchez-Garrido, Adri{\'a}n and Shir, Ruth and Sonner, Julian",
    title = "{Krylov Complexity}",
    eprint = "2507.06286",
    archivePrefix = "arXiv",
    primaryClass = "hep-th",
    reportNumber = "CERN-TH-2025-128",
    month = "7",
    year = "2025"
}

@article{Berkooz:2018jqr,
    author = "Berkooz, Micha and Isachenkov, Mikhail and Narovlansky, Vladimir and Torrents, Genis",
    title = "{Towards a full solution of the large N double-scaled SYK model}",
    eprint = "1811.02584",
    archivePrefix = "arXiv",
    primaryClass = "hep-th",
    doi = "10.1007/JHEP03(2019)079",
    journal = "JHEP",
    volume = "03",
    pages = "079",
    year = "2019"
}

@article{Berkooz:2018qkz,
    author = "Berkooz, Micha and Narayan, Prithvi and Simon, Joan",
    title = "{Chord diagrams, exact correlators in spin glasses and black hole bulk reconstruction}",
    eprint = "1806.04380",
    archivePrefix = "arXiv",
    primaryClass = "hep-th",
    doi = "10.1007/JHEP08(2018)192",
    journal = "JHEP",
    volume = "08",
    pages = "192",
    year = "2018"
}

@article{Lin:2023trc,
    author = "Lin, Henry W. and Stanford, Douglas",
    title = "{A symmetry algebra in double-scaled SYK}",
    eprint = "2307.15725",
    archivePrefix = "arXiv",
    primaryClass = "hep-th",
    doi = "10.21468/SciPostPhys.15.6.234",
    journal = "SciPost Phys.",
    volume = "15",
    number = "6",
    pages = "234",
    year = "2023"
}

@article{Lin:2022rbf,
    author = "Lin, Henry W.",
    title = "{The bulk Hilbert space of double scaled SYK}",
    eprint = "2208.07032",
    archivePrefix = "arXiv",
    primaryClass = "hep-th",
    doi = "10.1007/JHEP11(2022)060",
    journal = "JHEP",
    volume = "11",
    pages = "060",
    year = "2022"
}

@article{Caputa:2021sib,
    author = "Caputa, Pawel and Magan, Javier M. and Patramanis, Dimitrios",
    title = "{Geometry of Krylov complexity}",
    eprint = "2109.03824",
    archivePrefix = "arXiv",
    primaryClass = "hep-th",
    doi = "10.1103/PhysRevResearch.4.013041",
    journal = "Phys. Rev. Res.",
    volume = "4",
    number = "1",
    pages = "013041",
    year = "2022"
}

@article{Barbon:2019wsy,
    author = "Barb{\'o}n, J. L. F. and Rabinovici, E. and Shir, R. and Sinha, R.",
    title = "{On The Evolution Of Operator Complexity Beyond Scrambling}",
    eprint = "1907.05393",
    archivePrefix = "arXiv",
    primaryClass = "hep-th",
    reportNumber = "IFT-UAM/CSIC-19-98",
    doi = "10.1007/JHEP10(2019)264",
    journal = "JHEP",
    volume = "10",
    pages = "264",
    year = "2019"
}

@article{Susskind:1994vu,
    author = "Susskind, Leonard",
    title = "{The World as a hologram}",
    eprint = "hep-th/9409089",
    archivePrefix = "arXiv",
    reportNumber = "SU-ITP-94-33",
    doi = "10.1063/1.531249",
    journal = "J. Math. Phys.",
    volume = "36",
    pages = "6377--6396",
    year = "1995"
}

@article{tHooft:1993dmi,
    author = "'t Hooft, Gerard",
    title = "{Dimensional reduction in quantum gravity}",
    eprint = "gr-qc/9310026",
    archivePrefix = "arXiv",
    reportNumber = "THU-93-26",
    journal = "Conf. Proc. C",
    volume = "930308",
    pages = "284--296",
    year = "1993"
}

@article{Maldacena:1997re,
    author = "Maldacena, Juan Martin",
    title = "{The Large $N$ limit of superconformal field theories and supergravity}",
    eprint = "hep-th/9711200",
    archivePrefix = "arXiv",
    reportNumber = "HUTP-97-A097, HUTP-98-A097",
    doi = "10.4310/ATMP.1998.v2.n2.a1",
    journal = "Adv. Theor. Math. Phys.",
    volume = "2",
    pages = "231--252",
    year = "1998"
}

@article{Jackiw:1984je,
    author = "Jackiw, R.",
    editor = "Baier, R. and Satz, H.",
    title = "{Lower Dimensional Gravity}",
    reportNumber = "MIT-CTP-1203",
    doi = "10.1016/0550-3213(85)90448-1",
    journal = "Nucl. Phys. B",
    volume = "252",
    pages = "343--356",
    year = "1985"
}

@article{Teitelboim:1983ux,
    author = "Teitelboim, C.",
    title = "{Gravitation and Hamiltonian Structure in Two Space-Time Dimensions}",
    doi = "10.1016/0370-2693(83)90012-6",
    journal = "Phys. Lett. B",
    volume = "126",
    pages = "41--45",
    year = "1983"
}

@misc{KitaevEntangled2015,
  author       = {Kitaev, Alexei},
  title        = {Entangled 2015: Kitaev Lecture},
  howpublished = {\url{https://online.kitp.ucsb.edu/online/entangled15/kitaev/}},
  year         = {2015},
  note         = {\texttt{KITP UCSB Online}}
}

@article{Sachdev:1992fk,
    author = "Sachdev, Subir and Ye, Jinwu",
    title = "{Gapless spin fluid ground state in a random, quantum Heisenberg magnet}",
    eprint = "cond-mat/9212030",
    archivePrefix = "arXiv",
    reportNumber = "PRINT-93-0077",
    doi = "10.1103/PhysRevLett.70.3339",
    journal = "Phys. Rev. Lett.",
    volume = "70",
    pages = "3339",
    year = "1993"
}

@article{Kitaev:2017awl,
    author = "Kitaev, Alexei and Suh, S. Josephine",
    title = "{The soft mode in the Sachdev-Ye-Kitaev model and its gravity dual}",
    eprint = "1711.08467",
    archivePrefix = "arXiv",
    primaryClass = "hep-th",
    doi = "10.1007/JHEP05(2018)183",
    journal = "JHEP",
    volume = "05",
    pages = "183",
    year = "2018"
}

@article{Cotler:2016fpe,
    author = "Cotler, Jordan S. and Gur-Ari, Guy and Hanada, Masanori and Polchinski, Joseph and Saad, Phil and Shenker, Stephen H. and Stanford, Douglas and Streicher, Alexandre and Tezuka, Masaki",
    title = "{Black Holes and Random Matrices}",
    eprint = "1611.04650",
    archivePrefix = "arXiv",
    primaryClass = "hep-th",
    reportNumber = "SU-ITP-16-19, SU-ITP-16/19, YITP-16-124",
    doi = "10.1007/JHEP05(2017)118",
    journal = "JHEP",
    volume = "05",
    pages = "118",
    year = "2017",
    note = "[Erratum: JHEP 09, 002 (2018)]"
}

@article{Brown:2015lvg,
    author = "Brown, Adam R. and Roberts, Daniel A. and Susskind, Leonard and Swingle, Brian and Zhao, Ying",
    title = "{Complexity, action, and black holes}",
    eprint = "1512.04993",
    archivePrefix = "arXiv",
    primaryClass = "hep-th",
    doi = "10.1103/PhysRevD.93.086006",
    journal = "Phys. Rev. D",
    volume = "93",
    number = "8",
    pages = "086006",
    year = "2016"
}

@article{Susskind:2014rva,
    author = "Susskind, Leonard",
    title = "{Computational Complexity and Black Hole Horizons}",
    eprint = "1403.5695",
    archivePrefix = "arXiv",
    primaryClass = "hep-th",
    doi = "10.1002/prop.201500092",
    journal = "Fortsch. Phys.",
    volume = "64",
    pages = "24--43",
    year = "2016",
    note = "[Addendum: Fortsch.Phys. 64, 44--48 (2016)]"
}

@article{Susskind:2014Comp,
author = {Susskind, Leonard},
title = {Computational complexity and black hole horizons},
journal = {Fortschritte der Physik},
volume = {64},
number = {1},
pages = {24-43},
doi = {https://doi.org/10.1002/prop.201500092},
url = {https://onlinelibrary.wiley.com/doi/abs/10.1002/prop.201500092},
eprint = {https://onlinelibrary.wiley.com/doi/pdf/10.1002/prop.201500092},
year = {2016}
}

@article{Harlow:2018tqv,
    author = "Harlow, Daniel and Jafferis, Daniel",
    title = "{The Factorization Problem in Jackiw-Teitelboim Gravity}",
    eprint = "1804.01081",
    archivePrefix = "arXiv",
    primaryClass = "hep-th",
    doi = "10.1007/JHEP02(2020)177",
    journal = "JHEP",
    volume = "02",
    pages = "177",
    year = "2020"
}

@article{Maldacena:2016hyu,
    author = "Maldacena, Juan and Stanford, Douglas",
    title = "{Remarks on the Sachdev-Ye-Kitaev model}",
    eprint = "1604.07818",
    archivePrefix = "arXiv",
    primaryClass = "hep-th",
    doi = "10.1103/PhysRevD.94.106002",
    journal = "Phys. Rev. D",
    volume = "94",
    number = "10",
    pages = "106002",
    year = "2016"
}

@article{Parker:2018yvk,
    author = "Parker, Daniel E. and Cao, Xiangyu and Avdoshkin, Alexander and Scaffidi, Thomas and Altman, Ehud",
    title = "{A Universal Operator Growth Hypothesis}",
    eprint = "1812.08657",
    archivePrefix = "arXiv",
    primaryClass = "cond-mat.stat-mech",
    doi = "10.1103/PhysRevX.9.041017",
    journal = "Phys. Rev. X",
    volume = "9",
    number = "4",
    pages = "041017",
    year = "2019"
}

@article{Balasubramanian:2022tpr,
    author = "Balasubramanian, Vijay and Caputa, Pawel and Magan, Javier M. and Wu, Qingyue",
    title = "{Quantum chaos and the complexity of spread of states}",
    eprint = "2202.06957",
    archivePrefix = "arXiv",
    primaryClass = "hep-th",
    doi = "10.1103/PhysRevD.106.046007",
    journal = "Phys. Rev. D",
    volume = "106",
    number = "4",
    pages = "046007",
    year = "2022"
}

@article{Rabinovici:2020ryf,
    author = "Rabinovici, E. and S{\'a}nchez-Garrido, A. and Shir, R. and Sonner, J.",
    title = "{Operator complexity: a journey to the edge of Krylov space}",
    eprint = "2009.01862",
    archivePrefix = "arXiv",
    primaryClass = "hep-th",
    doi = "10.1007/JHEP06(2021)062",
    journal = "JHEP",
    volume = "06",
    pages = "062",
    year = "2021"
}

@article{Arik:1973vg,
    author = "Arik, M. and Coon, Darryl D.",
    title = "{Hilbert Spaces of Analytic Functions and Generalized Coherent States}",
    reportNumber = "PITT-116",
    doi = "10.1063/1.522937",
    journal = "J. Math. Phys.",
    volume = "17",
    pages = "524",
    year = "1976"
}

@article{Lin:2022zxd,
    author = "Lin, Henry W. and Maldacena, Juan and Rozenberg, Liza and Shan, Jieru",
    title = "{Looking at supersymmetric black holes for a very long time}",
    eprint = "2207.00408",
    archivePrefix = "arXiv",
    primaryClass = "hep-th",
    doi = "10.21468/SciPostPhys.14.5.128",
    journal = "SciPost Phys.",
    volume = "14",
    number = "5",
    pages = "128",
    year = "2023"
}

@article{Bagrets:2016cdf,
    author = "Bagrets, Dmitry and Altland, Alexander and Kamenev, Alex",
    editor = "Unno, Yoshinobu and Ohsugi, Takashi and Hou, Suen and Sadrozinski, Hartmut F. -W. and Lou, Xinchou and Zhu, Hongbo and Ouyang, Qun",
    title = "{Sachdev{\textendash}Ye{\textendash}Kitaev model as Liouville quantum mechanics}",
    eprint = "1607.00694",
    archivePrefix = "arXiv",
    primaryClass = "cond-mat.str-el",
    doi = "10.1016/j.nuclphysb.2016.08.002",
    journal = "Nucl. Phys. B",
    volume = "911",
    pages = "191--205",
    year = "2016"
}

@article{Fu:2016vas,
    author = "Fu, Wenbo and Gaiotto, Davide and Maldacena, Juan and Sachdev, Subir",
    title = "{Supersymmetric Sachdev-Ye-Kitaev models}",
    eprint = "1610.08917",
    archivePrefix = "arXiv",
    primaryClass = "hep-th",
    doi = "10.1103/PhysRevD.95.026009",
    journal = "Phys. Rev. D",
    volume = "95",
    number = "2",
    pages = "026009",
    year = "2017",
    note = "[Addendum: Phys.Rev.D 95, 069904 (2017)]"
}

@article{Chapman:2024pdw,
    author = "Chapman, Shira and Demulder, Saskia and Galante, Dami{\'a}n A. and Sheorey, Sameer U. and Shoval, Osher",
    title = "{Krylov complexity and chaos in deformed Sachdev-Ye-Kitaev models}",
    eprint = "2407.09604",
    archivePrefix = "arXiv",
    primaryClass = "hep-th",
    doi = "10.1103/PhysRevB.111.035141",
    journal = "Phys. Rev. B",
    volume = "111",
    number = "3",
    pages = "035141",
    year = "2025"
}

@article{Chryssanthacopoulos:2025xyn,
    author = "Chryssanthacopoulos, James and Vegh, David",
    title = "{Krylov Complexity of Supersymmetric SYK Models}",
    eprint = "2511.20769",
    archivePrefix = "arXiv",
    primaryClass = "hep-th",
    reportNumber = "QMUL-PH-25-33",
    month = "11",
    year = "2025"
}

@article{Narayan:2023wlk,
    author = "Narayan, Prithvi and S, Swathi T.",
    title = "{SYK Model with global symmetries in the double scaling limit}",
    eprint = "2302.11882",
    archivePrefix = "arXiv",
    primaryClass = "hep-th",
    doi = "10.1007/JHEP05(2023)083",
    journal = "JHEP",
    volume = "05",
    pages = "083",
    year = "2023"
}

@article{Gubankova:2025gbx,
    author = "Gubankova, Elena and Sachdev, Subir and Tarnopolsky, Grigory",
    title = "{Scaling limits of complex Sachdev-Ye-Kitaev models and holographic geometry}",
    eprint = "2512.05294",
    archivePrefix = "arXiv",
    primaryClass = "hep-th",
    month = "12",
    year = "2025"
}

@article{Nandy:2024evd,
    author = "Nandy, Pratik and Matsoukas-Roubeas, Apollonas S. and Mart{\'\i}nez-Azcona, Pablo and Dymarsky, Anatoly and del Campo, Adolfo",
    title = "{Quantum dynamics in Krylov space: Methods and applications}",
    eprint = "2405.09628",
    archivePrefix = "arXiv",
    primaryClass = "quant-ph",
    reportNumber = "RIKEN-iTHEMS-Report-24",
    doi = "10.1016/j.physrep.2025.05.001",
    journal = "Phys. Rept.",
    volume = "1125-1128",
    pages = "1--82",
    year = "2025"
}

@article{Aguilar-Gutierrez:2025sqh,
    author = "Aguilar-Gutierrez, Sergio E.",
    title = "{Evolution With(out) Time: Relational Holography {\&} BPS Complexity Growth in $\mathcal{N}=2$ Double-Scaled SYK}",
    eprint = "2510.11777",
    archivePrefix = "arXiv",
    primaryClass = "hep-th",
    month = "10",
    year = "2025"
}

@article{Blommaert:2024ymv,
    author = "Blommaert, Andreas and Mertens, Thomas G. and Papalini, Jacopo",
    title = "{The dilaton gravity hologram of double-scaled SYK}",
    eprint = "2404.03535",
    archivePrefix = "arXiv",
    primaryClass = "hep-th",
    doi = "10.1007/JHEP06(2025)050",
    journal = "JHEP",
    volume = "06",
    pages = "050",
    year = "2025"
}

@article{Gaikwad:2018dfc,
    author = "Gaikwad, Adwait and Joshi, Lata Kh and Mandal, Gautam and Wadia, Spenta R.",
    title = "{Holographic dual to charged SYK from 3D Gravity and Chern-Simons}",
    eprint = "1802.07746",
    archivePrefix = "arXiv",
    primaryClass = "hep-th",
    doi = "10.1007/JHEP02(2020)033",
    journal = "JHEP",
    volume = "02",
    pages = "033",
    year = "2020"
}

@article{Mertens:2019tcm,
    author = "Mertens, Thomas G. and Turiaci, Gustavo J.",
    title = "{Defects in Jackiw-Teitelboim Quantum Gravity}",
    eprint = "1904.05228",
    archivePrefix = "arXiv",
    primaryClass = "hep-th",
    doi = "10.1007/JHEP08(2019)127",
    journal = "JHEP",
    volume = "08",
    pages = "127",
    year = "2019"
}
\end{document}